\RequirePackage{fix-cm}
\documentclass{svjour3}                     \smartqed  \usepackage{graphicx}
\usepackage{array}
\usepackage{amsmath}
\usepackage{amssymb}
\usepackage{amsfonts}
\usepackage{bm}
\usepackage{xcolor}
\usepackage{multirow,array}
\usepackage{comment}
\usepackage{url} 
\usepackage{tikz}
\usepackage[normalem]{ulem}
\usetikzlibrary{matrix}

\definecolor{coltwo}{HTML}{bdcfed}
\definecolor{colthree}{HTML}{e2f7ff}

\tikzset{ 
  table/.style={
    matrix of nodes,
    row sep=-\pgflinewidth,
    column sep=-\pgflinewidth,
    nodes={rectangle,text width=3.5cm,align=center, shading = axis,rectangle, left color=coltwo!60, right color=coltwo!5,shading angle=90},
    text depth=2.5ex,
    text height=2.5ex,
    nodes in empty cells
  }
}

\newcommand{\cbox}[1]{\parbox[t]{2.8cm}{\centering #1}}

\DeclareSymbolFont{bbold}{U}{bbold}{m}{n}
\DeclareSymbolFontAlphabet{\mathbbold}{bbold}

\providecommand{\patrick}[1]{\textcolor{magenta}{PM: {#1}} }
\providecommand{\dmr}[1]{\textcolor{teal}{DMR: {#1} }}

\DeclareMathOperator*{\E}{\mathbb{E}}

\newtheorem{Theo}{Theorem}

\newtheorem{Def}{Definition}

\begin{document}

\title{Multi-Objective Multi-Agent Decision Making:\\ A Utility-based Analysis and Survey}

\titlerunning{Multi-Objective Multi-Agent Decision Making}        

\author{Roxana R\u{a}dulescu         \and
        Patrick Mannion\and
        Diederik M.\ Roijers\and
        Ann Now\'{e}
}

\institute{Roxana R\u{a}dulescu \at
              Artificial Intelligence Lab \\
              Vrije Universiteit Brussel\\
              Belgium\\
              \email{roxana.radulescu@vub.be}
           \and
           Patrick Mannion \at
              School of Computer Science\\
              National University of Ireland Galway\\
              Ireland\\
              \email{mannion.patrick@gmail.com}
            \and
              Diederik M.\ Roijers \at
              Computational Intelligence\\
              Vrije Universiteit Amsterdam\\
              The Netherlands\\
              \email{d.m.roijers@vu.nl}
            \and
              Ann Now\'{e} \at
              Artificial Intelligence Lab\\
              Vrije Universiteit Brussel\\
              Belgium\\
              \email{ann.nowe@vub.be}
}


\maketitle

\begin{abstract}
The majority of multi-agent system (MAS) implementations aim to optimise agents' policies with respect to a single objective, despite the fact that many real-world problem domains are inherently multi-objective in nature. Multi-objective multi-agent systems (MOMAS) explicitly consider the possible trade-offs between conflicting objective functions.
We argue that, in MOMAS, such compromises should be analysed on the basis of the utility that these compromises have for the users of a system. As is standard in multi-objective optimisation, we model the user utility using utility functions that map value or return vectors to scalar values. This approach naturally leads to two different optimisation criteria: expected scalarised returns (ESR) and scalarised expected returns (SER). 
We develop a new taxonomy which classifies multi-objective multi-agent decision making settings, on the basis of the reward structures, and which and how utility functions are applied. This allows us to offer a structured view of the field, to clearly delineate the current state-of-the-art in multi-objective multi-agent decision making approaches and to identify promising directions for future research.
Starting from the execution phase, in which the selected policies are applied and the utility for the users is attained, we analyse which solution concepts apply to the different settings in our taxonomy. Furthermore, we define and discuss these solution concepts under both ESR and SER optimisation criteria. 
We conclude with a summary of our main findings and a discussion of many promising future research directions in multi-objective multi-agent systems.

\end{abstract}

\section{Introduction}
\label{sec:Introduction}

A multi-agent system (MAS) features multiple agents deployed into a common environment. This is an inherently distributed paradigm, which benefits from scalability (agents can be added as required) and fault tolerance (the failure of any one agent does not imply failure of the whole system). The agents within a MAS may act cooperatively, competitively, or may exhibit a mixture of these behaviours \cite{vlassis2007concise,Wooldridge01}.

The majority of MAS implementations aim to optimise agent's policies with respect to a single objective, despite the fact that many real world problems are inherently multi-objective in nature.
Single-objective approaches seek to find a single policy to a problem, whereas in reality a system may have multiple possibly conflicting objectives. 
Multi-objective optimisation (MOO) \cite{deb2014multi} approaches consider these possibly conflicting objectives explicitly. In multi-objective multi-agent systems (MOMAS) the reward signal for each agent is a vector, where each component represents the performance on a different objective. By taking a multi-objective perspective on decision making problems, complex trade-offs can be managed; e.g., when selecting energy sources for electricity generation, there is an inherent trade-off between using cheap sources of energy which damage the environment, versus using renewable energy sources which are more expensive but better for the environment. 
Such trade-offs appear in a wide range of domains such as urban transportation, aviation, management of natural resources and robotics; these are all domains where multi-objective multi-agent approaches could confer huge benefits.

Compromises between competing objectives should be made on the basis of the utility that these compromises have for the users. In other words, if we can define a utility function that maps the vector value of a compromise solution to a scalar utility -- called a \textit{utility} or \textit{scalarisation} function -- then we can derive what to optimise \cite{roijers2017multi}, and how to measure the quality of solutions \cite{zintgraf2015quality}. In some rare cases, we might even be able to apply the utility function a priori, and try to solve the decision problem as a single-objective problem. However, as is known from single-agent multi-objective decision making \cite{roijers2013survey}, it is often impossible, undesirable, or unfeasible to perform such a priori scalarisation. For example, if the utility function is non-linear, this typically renders a priori scalarisation and subsequent single-objective solution methods intractable. Moreover, while trying to find compromise policies, i.e., while the agents are planning or learning, the utility function is often unknown or uncertain. In such cases, it is often desirable to construct a so-called \emph{coverage set}, a set of solutions that has at least one optimal policy with respect to every possible utility function that a user might have. 

The utility-based approach naturally leads to two different optimisation criteria for agents in a MOMAS: expected scalarised returns (ESR) and scalarised expected returns (SER). In the former, the users derive their utility from single roll-outs of the policy, while in the latter, the utility is derived from the expected outcomes, i.e., the mean over multiple roll-outs. To date, the differences between the SER and ESR approaches have received little attention in multi-agent settings, despite having received some attention in single-agent settings (see e.g. \cite{roijers2018multi,roijers2013survey}). Consequently, the implications of choosing either ESR or SER as the optimisation criterion for a MOMAS are currently not well-understood.

In single-agent or fully cooperative multi-agent settings \cite{roijers2017multi}, it is typically assumed that there is one, possibly unknown, utility function that determines the possibly unknown preferences of the users, and that the users are interested in the utility of the expected vector-valued returns (i.e., the SER optimality criterion). In the execution phase all agents will ultimately pursue the best utility with respect to this single function. Therefore, the coverage set can be derived from everything that is known about this utility function, and the types of policies allowed. For example, for deterministic stationary policies and possibly non-linear utility functions, a coverage set is a so-called Pareto front of deterministic stationary policies. That is the set of policies that are not Pareto-dominated, i.e., for which there is no other deterministic stationary policy that has a better or equal value for all objectives and is better in at least one objective. A Pareto-undominated policy is also called Pareto-optimal. Another well-known coverage set is a \emph{convex coverage set (CCS)}, which is a coverage set with respect to all possible linear utility functions. Incidentally, a Pareto-coverage set for stochastic policies can be constructed from a CCS of deterministic stationary policies \cite{vamplew2009constructing,roijers2013survey}. We discuss related work on single-objective and fully cooperative multi-agent systems in Section \ref{sec:single-agent}.

In multi-agent settings however, the situation can become much more complex. While in fully cooperative multi-agent systems, the agents are assumed to all receive the same team rewards, the individual reward vectors received by agents may be different in general multi-agent settings. We review various models with different assumptions regarding the reward functions, as well as observability and statefulness, in Section \ref{sec:Models}. Then, we consider settings where individual agents value objectives according to their own preferences, i.e., where each agent can have their own utility function. This has important consequences for what constitutes a solution set. We build a taxonomy of what constitutes a solution for a multi-objective multi-agent decision problem based on modelling assumptions, utility functions, and optimisation criteria, by analysing what happens at execution time in Section \ref{sec:execution}. We note that many of the different settings we identify in Section \ref{sec:execution} are under-explored in the current literature, and offer examples of decision problems that would merit further investigation for each part of our taxonomy. 

Using our taxonomy, we review the literature on multi-objective multi-agent decision problems in terms of optimal solution sets (Section \ref{sec:solutions}), and solution methods (Section \ref{sec:algorithmic}). Finally, we discuss what we consider the key open problems in this new and exciting field (Section \ref{sec:Future}).

\subsection{Motivating Example}
\label{sec:Example}

As a motivating example for adopting a multi-objective perspective on multi-agent decision making, we introduce a Multi-Objective Normal Form Game (MONFG) which is called the Commuting MONFG. By contrast to the usual Single-Objective Normal Form Game (SONFG) format, which is common in the literature, in a MONFG the agents receive payoffs in vector rather than scalar format after selecting their actions. This difference is illustrated in Tables \ref{table:General_SONFG} and \ref{table:General_MONFG}.

In the Commuting MONFG, two agents wish to commute from a common origin to the same destination. There are two transportation options available: travel by taxi or travel by train. If both agents choose the taxi option, they may split the cost equally between them. If they both choose to travel by train, they must each purchase their own ticket individually. If one chooses to travel by taxi and the other chooses to travel by train, they must also pay their own fares individually. A train ticket is cheaper than a taxi fare (even when agents share a taxi ride); however, the taxi journey takes less time than the train journey. The expected travel time and cost for each mode of transport is listed in Table \ref{table:CommutingMONFG_CostTime}. The individual or local vector payoffs for each agent are shown in Table \ref{table:CommutingMONFG_LocalPayoffs}. Note that the values in the matrix are negative as this is a minimisation problem for both objectives (commuters in general do not want to spend any additional time or money on their commute above what is necessary). 

From an utility-based perspective, each commuter will try to balance these conflicting objectives such that his/her derived utility is maximised. Each commuter can of course have a different utility function depending on how each objective is valued. Furthermore, depending on when each commuter evaluates the commute (e.g., on a monthly basis or after each trip), the two different optimisation criteria come into play: ESR or SER. 
For example, for some commuters travelling costs should be maintained within a certain budget every month, while still being on time at least 75\% of the time. This requires the use of the scalarised expected returns for the month. For other commuters the time component might be crucial and they cannot be late on any given day, thus imposing the expected scalarised return criterion.

  \begin{table}
  \centering
    \setlength{\extrarowheight}{2pt}
    \begin{tabular}{cc|c|c|}
      & \multicolumn{1}{c}{} & \multicolumn{2}{c}{Player $Y$}\\
      & \multicolumn{1}{c}{} & \multicolumn{1}{c}{$A$}  & \multicolumn{1}{c}{$B$} \\\cline{3-4}
      \multirow{2}*{Player $X$}  & $A$ & $(x_{A,A},y_{A,A})$ & $(x_{A,B},y_{A,B})$ \\\cline{3-4}
      & $B$ & $(x_{B,A},y_{B,A})$ & $(x_{B,B},y_{B,B})$ \\\cline{3-4}
    \end{tabular}
    \caption{General format of a Single-Objective Normal Form Game. $A$ and $B$ represent different actions which are available to the agents, and scalar payoffs $x$ and $y$ are given to each agent $X$ and $Y$ respectively depending on which combination of actions was selected.}
    \label{table:General_SONFG}
  \end{table}
  
  \begin{table}
  \centering
    \setlength{\extrarowheight}{2pt}
    \begin{tabular}{cc|c|c|}
      & \multicolumn{1}{c}{} & \multicolumn{2}{c}{Player $Y$}\\
      & \multicolumn{1}{c}{} & \multicolumn{1}{c}{$A$}  & \multicolumn{1}{c}{$B$} \\\cline{3-4}
      \multirow{2}*{Player $X$}  & $A$ & $(\mathbf{x}_{A,A},\mathbf{y}_{A,A})$ & $(\mathbf{x}_{A,B},\mathbf{y}_{A,B})$ \\\cline{3-4}
      & $B$ & $(\mathbf{x}_{B,A},\mathbf{y}_{B,A})$ & $(\mathbf{x}_{B,B},\mathbf{y}_{B,B})$ \\\cline{3-4}
    \end{tabular}
    \caption{General format of a Multi-Objective Normal Form Game. $A$ and $B$ represent different actions which are available to the agents, and vector payoffs $\mathbf{x}$ and $\mathbf{y}$ are given to each agent $X$ and $Y$ respectively depending on which combination of actions was selected.}
    \label{table:General_MONFG}
  \end{table}
  
  \begin{table}
  \centering
    \setlength{\extrarowheight}{2pt}
    \begin{tabular}{cc|c|c|}
      & \multicolumn{1}{c}{} & \multicolumn{2}{c}{}\\
      & \multicolumn{1}{c}{} & \multicolumn{1}{c}{cost}  & \multicolumn{1}{c}{time} \\\cline{3-4}
      \multirow{2}*{}  & taxi & $20$ & $10$ \\\cline{3-4}
      & train & $5$ & $30$ \\\cline{3-4}
    \end{tabular}    
    \caption{Cost and travel time for different modes of transport in the Commuting MONFG.}
    \label{table:CommutingMONFG_CostTime}
  \end{table}
  
  \begin{table}
  \centering
    \setlength{\extrarowheight}{2pt}
    \begin{tabular}{cc|c|c|}
      & \multicolumn{1}{c}{} & \multicolumn{2}{c}{Player $Y$}\\
      & \multicolumn{1}{c}{} & \multicolumn{1}{c}{taxi}  & \multicolumn{1}{c}{train} \\\cline{3-4}
      \multirow{2}*{Player $X$}  & taxi & $([-10,-10],[-10,-10])$ & $([-20,-10],[-5,-30])$ \\\cline{3-4}
      & train & $([-5,-30],[-20,-10])$ & $([-5,-30],[-5,-30])$ \\\cline{3-4}
    \end{tabular}    
    \caption{Individual (local) payoff matrix for the Commuting MONFG}
    \label{table:CommutingMONFG_LocalPayoffs}
  \end{table}

\section{Background}
\label{sec:Background}
Before we address the specifics of multi-objective multi-agent systems, and how to define optimal solutions for the Commuting MONFG in the previous section, we first introduce relevant background work on multi-agent decision theory, multi-objective decision making, optimisation criteria and utility functions which is necessary to understand the material covered later in this article.

\subsection{Multi-Agent Decision Theory}
\label{sec:madm}
Multi-agent systems appeared as a natural paradigm for modelling numerous real-world problems (e.g., health-care \cite{hurtado2018towards}, smart grid management \cite{moradi2016state,khan2017research}, traffic \cite{hamidi2018approach}, and Internet of Things \cite{Calvaresi2017}) as they lend themselves perfectly to the idea of large distributed systems.
They combine several disciplines ranging from artificial intelligence, software engineering, economics to social sciences \cite{Wooldridge01}.
We are mostly interested here in autonomous intelligent systems, where multiple agents are deployed in the same environment and are faced with a series of decision making problems.

Multi-agent decision making problems can often be modelled as a stochastic (or Markov) game (SG) \cite{shapley1953stochastic,vlassis2007concise}. A stochastic game can be formally defined as a tuple: $M = (S, \mathcal{A}, T, \mathcal{R})$, with $n\ge2$ agents, where:
\begin{itemize}
\item $S$ is the system state space
\item $\mathcal{A} = A_1 \times \dots \times A_n$ is the set of joint actions, $A_i$ is the action set of agent $i$
\item $T \colon S \times \mathcal{A} \times S \to \left[ 0, 1 \right]$ is a probabilistic transition function
\item $\mathcal{R}={R}_1 \times \dots \times {R}_n$ are the environment reward functions, where ${R}_i \colon S \times \mathcal{A} \times S \to \mathbb{R}$ is the reward function of agent $i$
\end{itemize} 

At every timestep, the environment emits a joint state $\bm{s}= \langle s_1, \dots, s_n\rangle$, where $\bm{s} \in S$. Depending on the system, the agents can fully observe this state, or can only observe a local view $s_i$. Notice that the reward received by an agent depends on the joint action taken by all the agents in the environment and not just on her own. 

However, the SG model is not the most general model. The stochastic game model can be further generalised to a partially observable stochastic game (POSG) \cite{hansen2004dynamic,wiggers2016structure} to include the possibility that the agents do not have full access to the environment state. In this case, each agent receives an observation and maintains a set of beliefs over possible states. Because the issue of partial observability is orthogonal to the existence of multiple objectives, but does make the model significantly more complex, we will restrict ourselves to fully observable models in this article. Note however that the solution concepts presented here generalise to partially observable environments as well. 

The behaviour of an agent is defined by its policy ${\pi}_i: S \times A_i \to \left[ 0, 1 \right]$, meaning that given a state, actions are selected according to a certain probability distribution. In the discounted infinite-horizon case, optimising $\pi_i$ is equivalent to maximising the expected discounted long-term reward:

\begin{equation}
V^{\pi_i} = \mathbb{E} \left[ \sum\limits^\infty_{t=0} \gamma^t r_{i,t} \:|\: \pi, \mu_0 \right]
\label{eqn:value_sosa}
\end{equation}
where $\pi$ is the joint policy of the agents acting in the environment, $\mu_0$ is the distribution over initial states $s_0$, $\gamma$ is the discount factor and $r_{i,t} = R_i(s_t, \mathbf{a}_{t}, s_{t+1})$ is the reward obtained by agent $i$ at timestep $t$, for the joint action $\mathbf{a}_{t}\in \mathcal{A}$, at state $s_t \in S$ and transitioning to the next state $s_{t+1} \in S$. 

Learning in multi-agent systems is considered a vital component, as environments are often characterised by high complexity and stochasticity, meaning that optimal behaviours are often impossible to achieve using pre-programmed approaches \cite{alonso2001learning}. However, we note that transitioning from single- to multi-agent learning is not straightforward. Building an intelligent distributed system is a notoriously difficult problem as it involves dealing with non-stationarity, limited resource sharing,  and often requires coordination or overcoming conflicting goals \cite{Sen:1999}. As a learning paradigm, we mainly consider reinforcement learning (RL) \cite{SuttonBarto1998}, but we will also discuss approaches from related fields such as game theory, planning or negotiation.

Multi-agent decision making is a multifaceted problem that can be explored through the lens of many fields and from different perspectives (e.g., system versus agent point of view). But perhaps the most important distinction we observe in multi-agent learning problems concerns the \emph{definition of the reward function}. Whether agents in a MAS will learn to act cooperatively or competitively depends directly on the reward function definitions. The literature usually distinguishes between three different settings:

\begin{itemize}
    \item \emph{cooperative}, where the reward function is the same for all agents: ${R}_1 = \dots = {R}_n = R$. Examples of this setting include domains where all agents work together to optimise the performance of a larger system, such as urban traffic control \cite{Mannion2016Experimental}, air traffic control \cite{chung2018multiagent} and electricity generator scheduling \cite{Mannion2016Multi}.
    \item \emph{competitive}, where any win for one agent implies a loss for another. Some competitive settings are zero-sum. Examples of this setting include fully competitive games such as Backgammon \cite{Tesauro1994TD}, Go \cite{silver2016mastering} and Starcraft 2 \cite{vinyals2017starcraft}.
    \item \emph{mixed}, where no restriction is imposed on the reward function definitions. Mixed games may incorporate elements of both cooperation and competition. Examples of this setting include games with opposing teams of agents, such as RoboCup soccer \cite{kitano1997robocup}.
\end{itemize}

This classification of multi-agent decision making problems is also reflected in the taxonomy we define in Section~\ref{sec:execution}.

Discrete, tabular representations are the simplest way for agents to store information which they have learned (e.g., policies, environment models, or action values in the case of model-free RL agents). When information is stored discretely, each additional feature tracked in the state leads to an exponential growth in the number of state-action pair values that must be stored \cite{SuttonBarto1998}. This problem is commonly referred to in the literature as the ``curse of dimensionality'', a term originally coined by Bellman \cite{Bellman1957Dynamic}. While this rarely occurs in simple environments, it may lead to an intractable learning task in complex real-world domains due to memory and/or computational constraints. Learning over a large state--action space is possible, but may take an unacceptably long time to learn useful policies.

Function approximation methods, such as artificial neural networks (ANNs), may be employed to represent policies, environment models or action values. These methods allows one to handle higher dimensional inputs, 
as well as allowing generalisation between similar observations and actions. Agents using function approximators can also potentially deal with continuous observation and/or action spaces. Recent advances in single and multi-agent RL make use of deep ANNs as function approximators; this emerging paradigm is known as deep reinforcement learning (DRL). For further information on recent deep MARL methods, the interested reader is referred to recent survey articles \cite{hernandez2018multiagent,nguyen2018deep}.

Another dimension to characterise multi-agent systems is represented by the degree of decentralisation. The planning/learning phase and the execution phase may be either (partially) centralised or fully decentralised. The paradigm of \emph{centralised training with decentralised execution} represents a middle ground between fully centralised and decentralised settings often used in cooperative or mixed settings \cite{foerster2016learning,foerster2017stabilising,lowe2017multi,radulescu2018deep,song2018multi}. The aim here is to enrich and aid the training/learning phase with extra information shared between the agents, however during the policy execution phase, the agents act in a fully decentralised manner.

\subsection{Multi-Objective Decision-Making}\label{sec:single-agent}
Single-objective decision making requires the existence of a single scalar reward function that agents can observe. The goal for the agents is then to find a policy that maximises the expected sum of these scalar rewards. However, most real-world problems do not adhere to this requirement. Specifically, there are typically multiple objectives that agents should care about. For example, consider the cost and time objectives of our transportation example in Section \ref{sec:Example}. 

\begin{figure}[!h]
    \centering
    \includegraphics[width=0.7\textwidth]{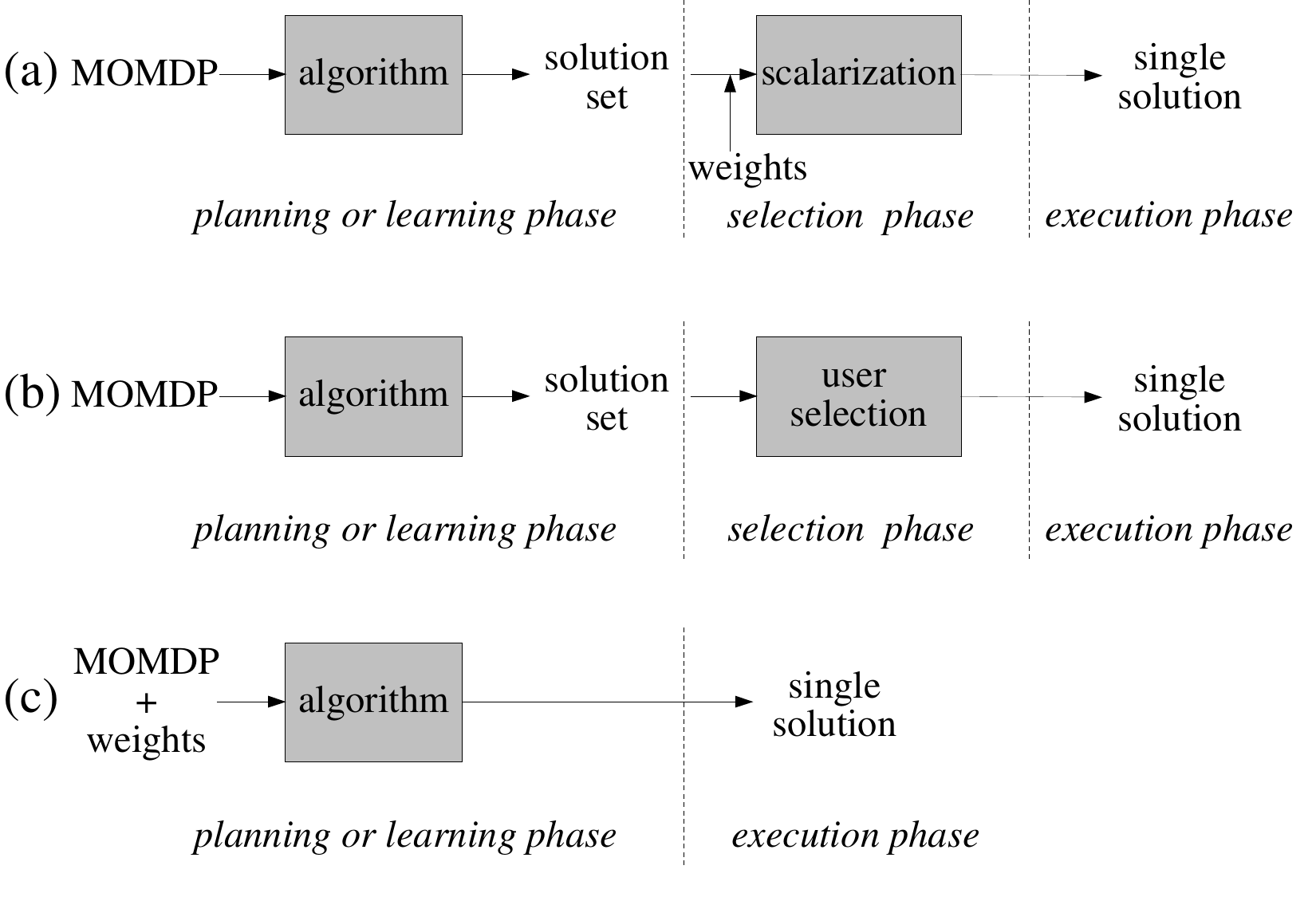}
\caption{Use-case scenarios for multi-objective decision making \cite{roijers2013survey}}
    \label{fig:scenariosSA}
\end{figure}

When there are multiple objectives, one might in some special cases still be able to use single-objective methods by using \emph{a priori scalarisation}. Specifically, if there exists a function that maps every possible outcome, i.e., a vector with policy values in each objective, which captures the exact preferences that the user(s) might have with respect to all these possible outcomes of the decision problem; this function is known a priori; and can be applied to the decision problem in such a way that the resulting single-objective problem remains tractable; then single-objective methods could still be used to solve the problem \cite{roijers2013survey}. However, often such a priori scalarisation is either impossible, infeasible or undesirable. 
Roijers et al. \cite{roijers2013survey} identify three use-case scenarios for multi-objective decision making, shown in Figure \ref{fig:scenariosSA}, in which this is indeed the case.  

In the \emph{unknown weights scenario}, or more precisely the \emph{unknown utility function scenario} (Figure \ref{fig:scenariosSA}a), a priori scalarsation is undesirable, as the utility that the user is able to get from the alternatives is too uncertain, or even unknown at the moment when planning or learning must occur. For example, when the objectives correspond to things that can be purchased or sold at an open market, but due to the complexity of the planning problem the prices can change significantly before planning or learning is complete. In such cases, it is desirable to compute a coverage set in order to be able to respond as quickly as possible whenever the available information about the market prices is updated. 
 
 In the \emph{decision support scenario} (Figure \ref{fig:scenariosSA}b), a priori scalarisation is infeasible or impossible, as a utility function that corresponds to the preferences of the user is never known explicitly. For example, consider a decision on the medical treatment of a serious illness. This decision problem has objectives such as maximising the probability of being cured and minimising the side effects. However, it is very difficult for a patient to articulate an exact utility function that makes all hypothetical trade-offs between these objectives a priori. In such cases it is therefore highly preferable to create a set containing the available possibly optimal alternatives, and present this set to the user. The decision support scenario thus proceeds almost identically to the unknown weights scenario.  The only difference is that in the selection
phase, the user selects a policy from the coverage set according to her arbitrary preferences, rather than explicit scalarization according to given weights. 
 
Finally, in the \emph{known weights scenario} or \emph{known utility function scenario}, a priori scalarisation would in principle be possible, as an exact utility function is available before planning or learning. However, even if this is indeed the case, it can still be undesirable to do so, because performing a priori scalarisation can lead to an intractable problem \cite{roijers2013survey}.

Key to all of these use cases is the notion of \emph{user utility}. Indeed, we argue that for any (multi-objective) decision making problem, the agent should always aim to maximise the user's utility. Specifically, following the work of Roijers et al. \cite{roijers2013survey}, we take the \emph{utility-based approach} to multi-objective decision making. In short, this means that the ultimate goal is to maximise \emph{user utility}, and that what constitutes a solution to a multi-objective decision problem should be derived from what is known about the user utility. User utility is characterised by a \emph{utility function} $u$ that maps vector-valued (expected) returns to a scalar value. We first discuss one optimisation criterion, SER, which will be discussed in the next section to give an impression of how an optimal solution set in multi-objective problems can be constructed. In SER, it is the value vector $\bm{V}^{\pi_i}$, i.e., the expected vector-valued return of policy $\pi_i$, that is projected to a scalar value:
\begin{equation}
V^{\pi_i} = u(\bm{V}^{\pi_i})
\end{equation}

\begin{figure}[ht]
\centering
\includegraphics[width=0.6\textwidth]{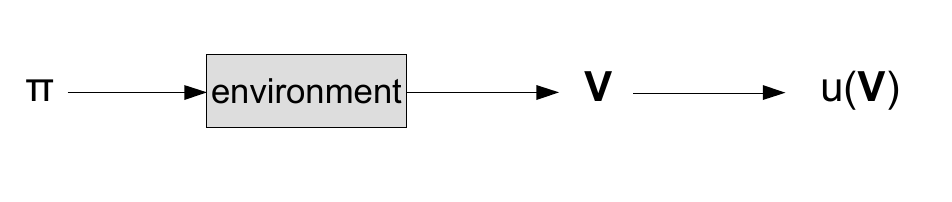}
\caption{The execution phase for single-agent multi-objective decision making.}
\label{fig:single-agent-exec}
\end{figure}

In order to derive the optimal solution set -- which is what a planning or learning algorithm should output -- one should start at the back of the use-case scenario, i.e., the execution phase, and work back, through the selection phase, until a specification of the optimal output of the algorithm is reached. 
As shown in Figure \ref{fig:single-agent-exec}, in single-agent multi-objective decision making, the execution phase is straight-forward.  
The agent uses its policy $\pi$ to interact with the environment, which leads to a value vector ${\bf V}^\pi$. Under SER (see next section) the utility function $u$ is applied to ${\bf V}^\pi$. This means that in the selection phase, the policy that maximises $u({\bf V}^\pi)$, must be available, which brings us to the selection phase. In a known weights scenario, this is trivial, as $u$ is known, so let us focus on the decision support and unknown weights scenarios. In both these scenarios, $u$ is (implicitly or explicitly) applied to a set of alternative value vectors, leading to the maximising policy from a set of alternatives to be chosen.\footnote{Note that we are assuming here that there is a small discrete set of alternatives, and that this maximisation can explicitly be computed in reasonable time. If this is not the case, for example if their set of alternatives is continuous, the user can be assisted in selecting a good policy using specific algorithms designed for the selection phase \cite{zintgraf2018ordered}. However, in such cases optimality can typically not be guaranteed.} Because in the unknown weights and decision support scenarios, $u$ is at least partially unknown when the agent needs to plan or learn, the planning or learning algorithm should output a set of alternative policies, that for every possible $u$ that a user might have (subject to what is already known at the beginning of the planning or learning phase), contains at least one optimal policy. This is called a \emph{coverage set} \cite{roijers2013survey}. 

In multi-agent settings, the execution phase is much less straight-forward. In fact, there are different settings, all with their own uses, that lead to very different schemas for the execution phase. Therefore, after discussing the various multi-agent multi-objective decision-theoretic problem settings in Section \ref{sec:Models}, we conduct a thorough analysis of the execution phase for multi-agent multi-objective decision making in Section \ref{sec:execution}.  Before going into different choices with respect to the execution phase in multi-agent settings however, we must first discuss another, perhaps even more fundamental choice: when to apply the utility function.  

\subsection{Multi-Objective Optimisation Criteria}
\label{sec:optimisation}
In the previous section, we have seen an example of how to construct an optimal solution for a given problem setting.
From a more general point of view, when agents consider multiple conflicting objectives, they should balance these in such a way that the user utility derived from the outcome of a decision problem is maximised. This is known as the utility-based approach \cite{roijers2013survey}. Following this approach, we assume that there exists a utility function that maps a vector with a value for each of the $d$ objectives to a scalar utility: $u \colon \mathbb{R}^d \to \mathbb{R}$.

We recall that the value function vector is defined similarly to Equation~\ref{eqn:value_sosa}, as the expected discounted long-term reward: $${\bf V}^{\pi} = \mathbb{E} \left[ \sum\limits^\infty_{t=0} \gamma^t {\bf r}_t \:|\: \pi, \mu_0 \right]$$ where $\mu_0$ is the distribution over initial states, $\pi$ is the agent's policy, $\gamma$ is the discount factor and ${\bf r}_t$ is the reward vector received for each of the objectives at timestep $t$.

When deciding what to optimise in a multi-objective decision making problem, we thus need to apply this utility function to the vector-valued outcomes of the decision problem in some way. 
There are two choices for how to do this  \cite{roijers2013survey,roijers2017multi}. Computing the expected value of the payoffs of a policy first and then applying the utility function, leads to the \emph{scalarised expected returns (SER)} optimisation criterion, i.e., 

\begin{equation}
    V_{u}^{\pi} = u\left(\mathbb{E} \left[ \sum\limits^\infty_{t=0} \gamma^t {\bf r}_t \:|\: \pi, \mu_0 \right]\right)
    \label{eqn:ser}
\end{equation}
\noindent where $V_{u}^{\pi}$ is the return derived by the agent from the vector ${\bf V}^{\pi}$. SER is employed in most of the multi-objective planning and reinforcement learning literature. Alternatively, the utility function can be applied before computing the expectation, leading to the \emph{expected scalarised returns (ESR)} optimisation criterion \cite{roijers2018multi}, i.e., 
\begin{equation}
    V_{u}^{\pi} = \mathbb{E} \left[ u\left( \sum\limits^\infty_{t=0} \gamma^t {\bf r}_t \right) \:|\: \pi, \mu_0 \right]
    \label{eqn:ser}
\end{equation}
Which of these criteria should be considered best depends on how we are interested in evaluating the outcome of a policy. SER is the correct criterion if we want to execute a policy multiple times, and it is the average return over multiple executions that determines the agent's utility. ESR is the correct formulation if the return of a single policy execution is what is important to the agent.

\subsection{Utility Functions}
\label{sec:utility}

Linear combinations are a widely used canonical example of a utility function:
\begin{equation}
    u(\mathbf{r}) = \sum\limits_{d \in D} w_d r_d
    \label{eqn:linear_utility}
\end{equation}
\noindent where $D$ is the set of objectives, $\mathbf{w}$ is a weight vector\footnote{A vector whose coordinates are all non-negative and sum up to 1.}, $w_d \in [0,1]$ is the weight for objective $d$ and $r_d$ is the component for objective $d$ of some reward vector $\mathbf{r}$. Interestingly, for such linear utility functions, there is no difference between SER and ESR, as the inner product with the weight vector distributes over the expectation. 

Non-linear, discontinuous utility functions may arise in the case where it is important for an agent to achieve a minimum payoff on one of the objectives; such a utility function may look like the following:
\begin{equation}
        u(\mathbf{r}) = \begin{cases}
        r_{t_d} & \text{if } r_d \geq t_d\\
        0 & \text{otherwise } 
        \end{cases}
        \label{eqn:nonlinear_utility_example}
\end{equation}
\noindent where $r_d$ represents the component of $\mathbf{r}$ for objective $d$, $t_d$ is the required threshold value for $d$, and $r_{t_d}$ is the reward for reaching the threshold value on $d$.
In general we are interested in the class of all monotonically increasing (possibly non-linear) utility functions. 
\begin{Def}\label{def:mono}
	A scalarization function $f$ is \emph{monotonically increasing} if:
	\begin{equation}\label{eq:strict}
	(\forall o,~V^\pi_o \geq V^{\pi'}_o) \Rightarrow u({\bf V}^\pi) > u({\bf V}^{\pi'}).
	\end{equation}
\end{Def}
This means that if for all objectives, the value for that objective under policy $\pi$ is greater than or equal than the value for that same objective under policy $\pi'$, then policy $\pi$ yields equal or higher utility than policy $\pi'$. This is a rather minimal assumption to make, as it translates to: we always want more of each objective. Non-linear utility functions may yield different optimal policies under SER and ESR. 

Utility functions may not always be known \emph{a priori} and/or may not be easy to define depending on the setting. For example, in the \emph{decision support scenario} \cite{roijers2013survey} it may not be possible for users to specify utility functions directly; instead users may be asked to provide their preferences by scoring or ranking different possible outcomes. After the preference elicitation process is complete, users' responses may then be used to model their utility functions \cite{zintgraf2018ordered}. 
\section{Modelling Multi-Objective Multi-Agent Settings}
\label{sec:Models}
In this section, we discuss how multi-objective problems with multiple agents can be modelled. We discuss the multi-objective stochastic game, and partially observable stochastic game models, as the most general models, and then show which additional assumptions can be made to arrive at more restricted models. 

\subsection{The Multi-Objective Stochastic Game Model}

As a framework for defining multi-objective multi-agent decision making settings we will use the Multi-Objective Stochastic Game (MOSG).
We formally define a MOSG as a tuple $M = (S, \mathcal{A}, T, \mathcal{R})$, with $n\ge2$ agents and $d\ge2$ objectives, where:
\begin{itemize}
\item $S$ state space
\item $\mathcal{A} = A_1 \times \dots \times A_n$ set of joint actions, $A_i$ is the action set of agent $i$
\item $T \colon S \times \mathcal{A} \times S \to \left[ 0, 1 \right]$ probabilistic transition function
\item $\mathcal{R}=\mathbf{R}_1 \times \dots \times \mathbf{R}_n$ reward functions, $\mathbf{R}_i \colon S \times \mathcal{A} \times S \to \mathbb{R}^d$ is the vectorial reward function of agent $i$ for each of the $d$ objectives 
\end{itemize} 

Furthermore, the same as in the stochastic game case, the MOSG can be extended to also incorporate partial observability. We can thus also define a multi-objective partially observable stochastic game (MOPOSG), where agents do not have access anymore to the full state of the environment. In this situation, agents receive observations from the environment and have to maintain beliefs over possible states. While, for the scope of this work, it is sufficient to consider the MOSG model, we will build our categorisations having the MOPOSG model as the most general case.

An agent behaves according to a policy ${\pi}_i: S \times A_i \to \left[ 0, 1 \right]$, meaning that given a state, actions are selected according to a certain probability distribution. Optimising $\pi_i$ is equivalent to maximising the expected discounted long-term reward:

\begin{equation}
\bm{V}^{\pi_i} = \mathbb{E} \left[ \sum\limits^\infty_{t=0} \gamma^t \mathbf{R}_i(s_t, \mathbf{a}_{t}, s_{t+1}) \:|\: \pi, \mu_0 \right]
\end{equation}
where $\pi$ is the joint policy of the agents acting in the environment, $\mu_0$ is the distribution over initial states $s_0$, $\gamma$ is the discount factor and $\mathbf{R}_i(s_t, \mathbf{a}_{t}, s_{t+1})$ is the vectorial reward obtained by agent $i$ for the joint action $\mathbf{a}_{t}\in \mathcal{A}$, at state $s_t \in S$. We also note that it is also possible to extend this framework to include the case in which the discount factor is different for each agent $i$ by replacing $\gamma$ with $\gamma_i$. 

The value function is also vectorial, $\bm{V}^{\pi_i} \in \mathbb{R}^d$. We consider that each agents also has an individual utility function $u_i$ to project $\bm{V}^{\pi_i}$ to a scalar value, as described in Section~\ref{sec:utility}.

Starting from this model, we will further develop a taxonomy in Section~\ref{sec:execution} focusing on the utility and reward axis. Furthermore, we show how the approaches found in the literature can be mapped to this model by limiting various dimensions such as states, observability, individual rewards, or utilities.

\subsection{Special Case Models}

\begin{figure}[!h]
    \centering
    \includegraphics[width=0.7\textwidth]{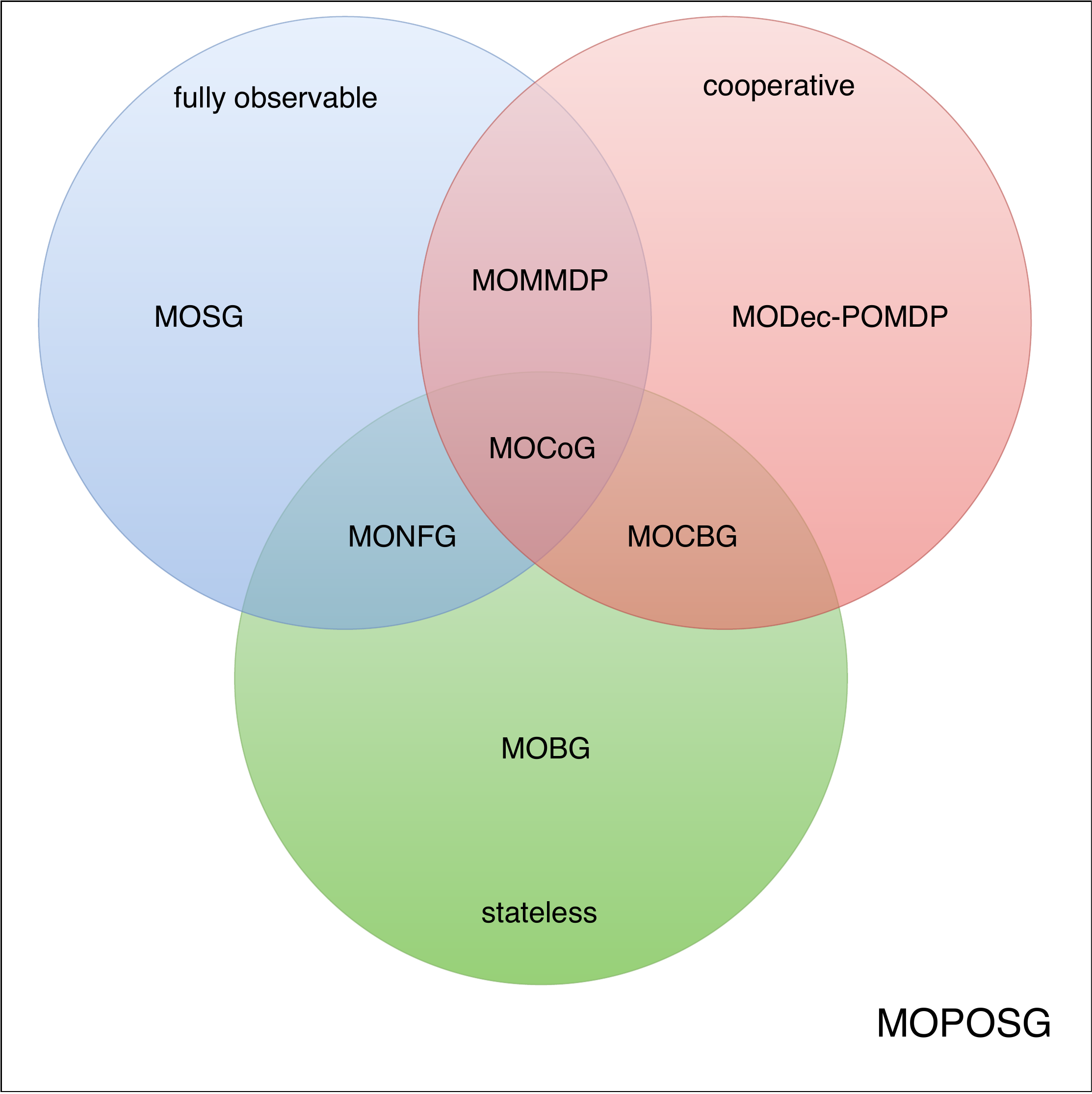}
\caption{The MOPOSG is a general model, which encompasses many other common decision making models. The abbreviations in the Venn diagram stand for: multi-objective partially observable stochastic game (MOPOSG), multi-objective stochastic game (MOSG), multi-objective decentralised partially observable Markov Decision Process (MODec-POMDP), multi-objective Bayesian game (MOBG), multi-objective multi-agent Markov Decision Process (MOMMDP), multi-objective normal form game (MONFG), multi-objective collaborative Bayesian game (MOCBG), and multi-objective coordination graph (MOCoG) }
    \label{fig:venn}
\end{figure}

The MOPOSG model is general enough to encompass a wide range of multi-objective multi-agent decision making settings; consequently, many prior decision making models may be viewed as special cases of a MOPOSG. By restricting certain degrees of freedom in the MOPOSG model, one can derive many commonly used decision making models from the single-agent and multi-agent literature, as well as the single-objective and multi-objective literature; e.g. by setting the number of agents $n=1$ and the number of objectives $d=1$ in a MOPOSG, one may obtain a traditional POMDP. Fig. \ref{fig:venn} outlines the relationship between the MOPOSG model and many other common multi-objective multi-agent decision making models. Table \ref{table:MOPOSG_restrictions} sumarises other common  decision making models, and outlines which degrees of freedom of the MOPOSG model must be restricted to derive each other model. We hope that readers will be able to use this as a reference, so they can easily identify ways in which problem settings and algorithms from the single-objective literature could be extended/reanalysed from a multi-objective perspective. Furthermore, it should be possible to easily spot methods developed specifically for multi-objective models which could be applied to the corresponding single-objective model.

\begin{table}[ht]
    \centering
    \begin{tabular}{l|l|l|c|c|c|c}
        & & Model & $d$ & $n$ & $|S|$ & observability \\
        \hline
        \multirow{12}*{multi-objective} & \multirow{9}*{multi-agent} & MOPOSG      &  &  &  &  \\
        &  & MOSG      &  &  &  & full \\
        &  & MODec-POMDP      &  &  &  &  \\
        &  & MOMMDP      &  &  &  & full \\
        &  & MOCoG      &  &  & 1 & full \\
        &  & MOBG      &  &  & 1 &  \\
        &  & MOCBG      &  &  & 1 &  \\
        &  & MONFG      &  & 2 & 1 & full \\
        &  & MOMG      &  & 2 & 1 & full \\
         \cline{2-7}
        & \multirow{3}*{single-agent} & MOPOMDP     &  & 1 &  &  \\
        &  & MOMDP     &  & 1 &  & full \\
        &  & MO Multi-armed bandit     &  & 1 & 1 & full \\
        \hline
        \multirow{12}*{single-objective} & \multirow{9}*{multi-agent} & POSG      &  &  &  &  \\
        &  & SG      & 1 &  &  & full\\
        &  & Dec-POMDP  & 1 &  &  & \\
        &  & MMDP      & 1 &  &  & full \\
        &  & CoG      & 1 &  & 1 & full \\
        &  & BG      & 1 &  & 1 &  \\
        &  & CBG      & 1 &  & 1 &  \\
        &  & NFG      & 1 & 2 & 1 & full  \\
        &  & MG      & 1 & 2 & 1 & full \\        
        \cline{2-7}
        & \multirow{3}*{single-agent} & POMDP      & 1 & 1 &  &  \\
        &  & MDP      & 1 & 1 &  & full \\
        &  & Multi-armed bandit      & 1 & 1 & 1 & full \\
    \end{tabular}
    \caption{Summary of which degrees of freedom must be restricted to derive common decision making models from the MOPOSG model. Here $d$ is the number of objectives, $n$ is the number of agents and $|S|$ is the size of the system state space. Blank cells indicate no restriction, whereas numeric values indicate a required parameter setting. (See Figure \ref{fig:venn} for a list of the abbreviations.)}
    \label{table:MOPOSG_restrictions}
\end{table}

\section{The Execution Phase}
\label{sec:execution}

As mentioned in Section \ref{sec:Models}, multi-objective multi-agent models are typically named according to assumptions about observability,  whether the problem is sequential, and the structure of the reward function. These are indeed important distinctions. However, following the utility-based approach \cite{roijers2013survey}, this information is not sufficient to determine what constitutes a solution for such a problem. Specifically, we should aim to optimise the utility of the user(s). In single-agent multi-objective problems, we can typically assume that at execution time we aim to optimise the utility of a single user with a single utility function\footnote{Or multiple users whose utility functions can be aggregated in an overall aggregated utility function}. The shape of the utility function, in conjunction with the allowed policy space, can be used to derive the optimal solution set that a multi-objective decision-theoretic algorithm should produce. 

In multi-agent settings, the situation is more complex than in single-agent settings. Particularly, each individual agent can represent one or more distinct users. In other words, the utility function may vary per agent. For example, assume we have a group of friends deciding where to go on holiday, who outsource the decision making to a group of agents (one agent per friend). The objectives they agree on are minimising costs, minimising the distance from the hotel to the beach, maximising the expected number of hours of sun, and maximising the number of museums and other points of cultural interest within a 20km radius. After a decision is reached, every friend will get the same (expected) returns vector. However, each friend may have a different utility for each possible vector---in fact this is the entire reason that this decision problem may be hard. Furthermore, it depends on which perspective we take, as the algorithm designers. In the example, we have taken the perspective of the individual users, but we could also take the perspective of an external observer that wants the outcome to be fair (for whatever definition of fair), i.e., wants to optimise some form of social welfare. 

\begin{figure}[!h]
    \centering
    \includegraphics[width=0.9\textwidth]{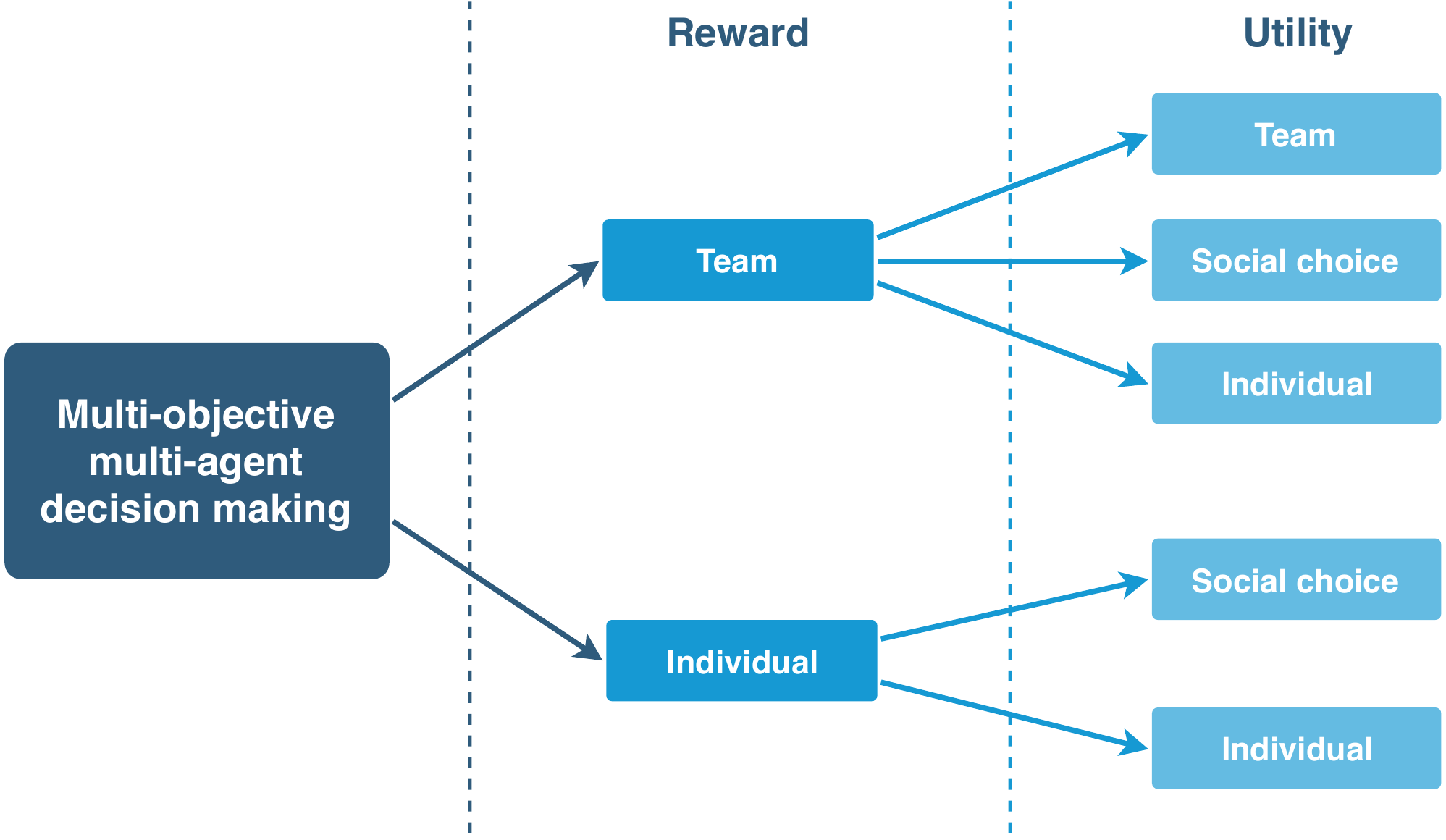}
\caption{Taxonomy of multi-objective multi-agent decision making settings.}\label{fig:taxonomy}
\end{figure}

We propose a taxonomy based on the \emph{reward} as well as the \emph{utility} functions. We distinguish between two types of reward functions: a \emph{team reward}, in which each agent receives the same value or return vector for executing the policy, and \emph{individual rewards} in which each agent receives a different value/return vector. Furthermore, we make a distinction in three types of \emph{utility}---more or less orthogonally to the types of rewards--- i.e., \emph{team utility}, which is what happens when all the agents serve the same interest, e.g., when they all work for a single company or are on the same football team; \emph{social choice utility}, when we are interested in optimising the overall social welfare across all agents; and \emph{individual utility}, which is what happens if each agent serves a different agenda and just tries to optimise for that. This results in the taxonomy provided in Figure \ref{fig:taxonomy}. We further note that the utility functions may be applied according to the ESR or SER criteria for every setting.

We note that in the taxonomy, the team reward and team utility setting could be translated to a single-agent setting, by flattening out the multi-agent aspect. Specifically, we could define a single agent that would control the actions of all other agents, i.e., one agent choosing its actions from the entire joint action space. As such, the solution concepts from the single-agent multi-objective literature apply \cite{roijers2013survey}. However, the problem can still be significantly harder than a single-agent problem, due to the size of the joint action space, as we discuss in  Section \ref{sec:trtu}. 

Furthermore, we note that the individual rewards with a team utility setting is not realistic; even if the utility function of all the individual agents would be the same (i.e., the agents have the same opinion about what is important), that would still lead to different individual utilities due to different input (expected) return vectors. Hence, even when the utility functions are identical, we treat these as \emph{individual utilities}. 

In the remainder of this section, we discuss each of the remaining settings in our taxonomy in more detail. In Section \ref{sec:solutions} we discuss the various solution concepts that apply to these settings (see Figure \ref{fig:taxonomy}).

\subsection{Team Reward}

First, we consider the top row of Figure \ref{fig:taxonomy}, \emph{team reward}. In this setting each agent receives the same reward vectors, ${\bf R}_1 = \dots = {\bf R}_n = {\bf R}$. As a result, the (expected) return vector is the same for each agent when a given joint policy is executed. This is for example the case in \emph{multi-objective multi-agent Markov decision processes (MOMMDPs)} \cite[Section 5.2.1]{roijersPhD}, as we discussed in the previous section. 

At first glance, this may appear to be a fully cooperative setting. However, this depends on how much the individual agents value their (expected) cumulative reward vectors, i.e., on the utility function of each agent. We distinguish between three cases: team utility, individual utility, and social choice with respect to individual utilities. 

\subsubsection{Team Reward Team Utility}
\label{sec:trtu}

Perhaps due to its relative simplicity, the most common case by far in the multi-objective multi-agent decision-theoretic planning and reinforcement learning literature is the team reward with team utility setting, i.e., all the agents together aim to strive for a single maximum utility, under SER,
\[
V^* = \max_{\pi} u(\bm{V}^{\pi}), 
\]
or under ESR: 
\[
V^* = \max_{\pi}  \mathbb{E}[u({\bm \rho})|\pi, \mu_0], 
\]
where $\bm{\rho} = \sum_{t=0}^\infty \gamma^t {\bf r_t} $. $u$ (including its parameterisation) may or may not be known to the agents. This is a truly fully cooperative setting. Therefore, the optimal solution sets, i.e., coverage sets, can be derived from the same information as in single-agent multi-objective settings (see Section \ref{sec:Background}), and the same types of solution methods apply. 

\begin{figure}[ht]
\centering
\includegraphics[width=0.6\textwidth]{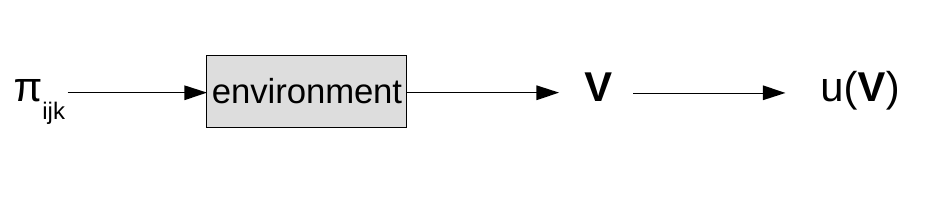}
\caption{The execution phase for the Team Reward Team Utility setting. This figure depicts the SER optimality criterion, where the expected values (i.e., the average over many executions of the policies) will be input to $u$. Under ESR the input to $u$ would be $\bm \rho$, i.e., the returns for an individual roll-out.}
\label{fig:trtu}
\end{figure}

Even though techniques similar to single-agent multi-objective settings can be used to solve multi-agent multi-objective settings, multi-agent multi-objective problems are much more complex than their single-agent counterparts. Specifically, the number of possible joint actions increases exponentially in the number of agents, leading to a much larger policy space. In turn, in cases where the utility function is unknown during planning or learning this leads to much larger coverage sets. 

To keep multi-objective multi-agent planning and reinforcement learning tractable in these settings, it is key to exploit so called \emph{loose couplings} \cite{Guestrin02,KokVlassis2004}, i.e., each agent's actions only directly affect a subset of the other agents. Loose couplings can be expressed using a factorised reward function. Such a factorised reward function can be visually represented as a graphical model known as a \emph{coordination graph} in the multi-agent literature. The single-shot setting -- the multi-objective coordination graph (MO-CoG) -- is one of the most well-studied models in the multi-objective multi-agent literature \cite[etc.]{delle2011bounded,dubus2009,dubus2009choquet,marinescu2009exploiting,marinescu2011efficient,roijers2015computing,roijers2015variational,roijers2017multi,Rollon06MOBE,rollonThesis,wilson2015}. Exploiting loose couplings also plays an important role in sequential multi-objective multi-agent settings \cite{scharpff2016solving,roijersPhD}. 

We discuss the solution concepts for this setting in Section \ref{sec:cs}.

\subsubsection{Team Reward Individual Utility}
\label{sec:triu}
\begin{figure}[ht]
\centering
\includegraphics[width=0.6\textwidth]{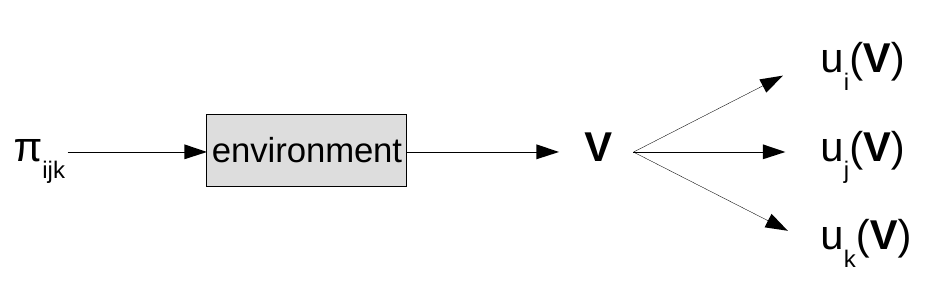}
\caption{The execution phase for the Team Reward Individual Utility setting. This figure depicts the SER optimality criterion, where the expected values (i.e., the average over many executions of the policies) will be input to $u$. Under ESR the input to $u$ would be $\bm \rho$, i.e., the returns for an individual roll-out.}
\label{fig:triu}
\end{figure}

When a group of agents receives a single shared reward vector, that does not mean that all agents value that reward equally. 
For example, imagine that you are playing a massive multiplayer online role playing game  (MMORPG), and you set out on a quest with teammates. You will play multiple quests with the same team, so you are interested in the expected returns rather than the returns of a single quest (SER). The expected value of doing a quest in terms of experience points, currency and gear is the same for each member of the team, but for different players each of these objectives may be more or less important. Therefore, even when the team gets team rewards for all quests, the members of the team may prefer different quests. This is because mathematically, each agent tries to optimise its own utility via the team value of a joint policy: 
\[
V^\pi_i = u_i(\bm{V}^{\pi}), 
\]
under SER, or, 
\[
V^\pi_i = \mathbb{E}[u_i({\bm \rho})|\pi, \mu_0], 
\]
under ESR, where $\pi$ is the joint team policy. This leads to the execution phase depicted in Figure \ref{fig:triu}.

The existence of individual utilities immediately poses a problem for the agents. Each agent can only control a small part of the joint policy, i.e., its own actions, and a lack of coordination may lead to a very bad policy for all agents. In other words, an agent cannot simply maximise its utility by changing its own policy without taking the policies, and policy changes, of the other agents into account. Therefore, in the selection phase -- immediately preceding the execution phase -- it is vitally important to coordinate, and agree on a joint policy.

There are two main ways to go about this. Firstly, let us view the game-theoretic perspective, in which we aim to find a joint policy that is in some sense \emph{stable}, i.e., agents do not have an incentive to deviate from the joint policy. Stable solutions come in many different levels of strictness \cite{chalkiadakis2011computational}, from core stability, to Nash equilibria, to individual stability. Particularly challenging in this respect is how to figure out what the individual preferences are. When agents do not or cannot divulge their individual utility functions a priori, for example because it would be hard or even impossible to specify this utility function exactly, algorithms that aim to find stable outcomes must learn about the individual utility functions of the agents to learn whether a joint policy is stable or not \cite{igarashi2017multi}. 

Finding a stable joint policy in the planning or learning phase may seem to mitigate the need for an extensive selection phase; as no agent will have an incentive to deviate from it, deviations should not happen. There are however two problems that could still arise. Firstly, if there are multiple possible stable solutions, the agents still need to agree on which of these to pick. Secondly, in repeated interaction settings, an agent could be spiteful, i.e., aim to be as disruptive to the elected stable solution as possible, in order to strengthen its hand the next time a stable solution must be selected. 

Secondly, there is the negotiation perspective, i.e., agents will try to hammer out a deal on which policy they will jointly execute. This has the advantage that even non-stable solutions---that may offer better utility for all agents than the stable ones---could be selected, as long as the agents are obligated to follow through.  For example, the Automated Negotiating Agents Competition (ANAC) \cite{jonker2017automated} considers three-agent negotiations in which agents negotiate about possible alternative outcomes. When each alternative is associated with its own vector corresponding to different objectives, the agents will know that some outcomes are Pareto-dominated, and should therefore be excluded from consideration, but for the solutions that are in the Pareto coverage set, different agents may have different preferences. In general, the outcome of such a negotiation should thus be a ``deal'' between the agents about which alternative joint policy from the coverage set to execute. 

Finally, we note that there is a special case of the team rewards and individual utilities, in which the number of objectives is equal to the number of agents, and the utility function of each agent would just be the value of the objective corresponding to that agent. This special case may seem identical to the single-objective multi-agent case with individual rewards, but there is in fact a significant difference. Specifically, it reflects the situation in which the agents can care about the rewards of the other agents, and can make (a priori or a posteriori) agreements on which division of rewards is admissible. In other words, it can be used to model various degrees of altruism. At a very minimum, the agents could all exclude Pareto-dominated solutions, leading to the situation in which agents always prefer to help the other agents to increase their rewards, as long as it does not cost them anything. A bit more drastically, the agents could agree to exclude a joint policy from consideration if another policy exists in which the total sum of the values for each objective/agent is at least the same, but more fairly distributed over the agents. This leads to the solution concept of Lorenz optimality \cite{golden2010infinite}, which we will discuss in Section~\ref{sec:lorenz}. 

\subsubsection{Team Reward and Social Welfare with Respect to Individual Utilities}
\label{sec:sw1}

In the individual rewards setting, it is hard to predict, let alone optimise for, utility. This is because the agents have different agendas, leading to complex behavioural dynamics, in which agents react to each other's behaviours. This process may not converge to stable solutions. Furthermore, the individual utility functions may not be common knowledge. 

\begin{figure}[ht]
\centering
\includegraphics[width=1\textwidth]{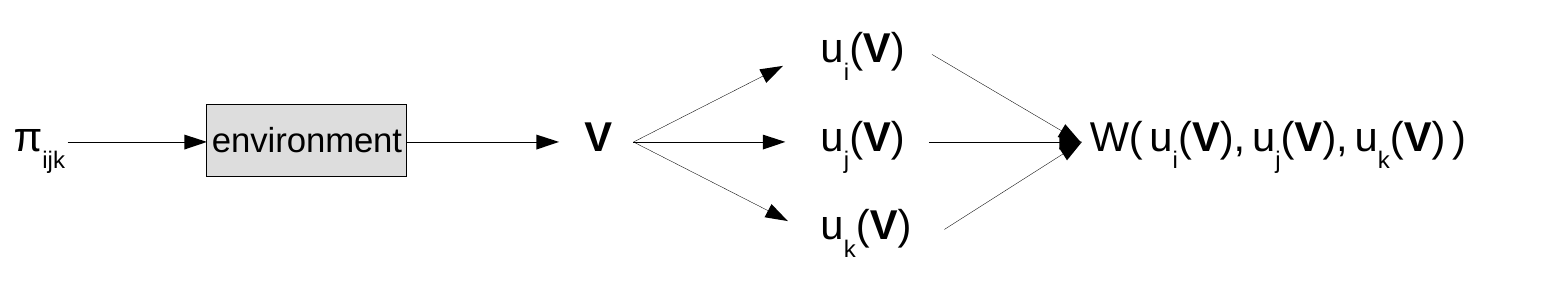}
\caption{The execution phase for the Team Reward and Social Welfare with Respect to Individual Utilities setting. Please note that the social welfare can depend \emph{both} on the utilities of the agents \emph{and} the value/return vector. This figure depicts the SER optimality criterion, where the expected values (i.e., the average over many executions of the policies) will be input to $u$. Under ESR the input to $u$ would be $\bm \rho$, i.e., the returns for an individual roll-out.}
\label{fig:triusw}
\end{figure}

A different perspective on this problem is to take a step back from the self-interested agents and optimising for their individual utilities, and instead look at what would be a \emph{desirable outcome}. For example, we can focus on what would be socially favourable by the agents. Once we have decided on what would be desirable, we can define \emph{social welfare} as a \emph{social choice function}, corresponding to the desirability of each outcome, and construct a system of payments that will make the joint policy converge to the desired outcome. This is known as mechanism design \cite[Ch.\ 6]{vlassis2007concise}. 

It is important to note that the social welfare function can depend both on the value or return vector, as well as the individual utilities of the agents, as illustrated in Fig. \ref{fig:triusw}. For example, in traffic, social welfare may depend on the pollution levels, as well as fairness between different vehicles in terms of their total expected time that they have to wait for traffic lights. 

In mechanism design, the challenge is to formulate the system of payments in such a way that the agents will be non-manipulable, i.e., do not have an incentive to lie about their preferences. If this succeeds, the agents will report their preferences truthfully, and from a planning perspective, the decision-problem becomes fully cooperative, i.e., aiming to collectively optimise the social choice function. 

For multi-objective decision problems, the social welfare perspective can for example be used by governments to control the parameters of tenders, to balance the different objectives for projects. For example, in a traffic network maintenance planning setting \cite{scharpff2013coordinating}, the balancing of traffic delays and costs can be made a posteriori, by computing a convex coverage set for a cooperative multi-objective multi-agent MDP \cite{Roijers2014Linear}, because a non-manipulable mechanism exists for every different weighting of the objectives. 
While mechanism design methods are very powerful, they do pose challenges. Specifically, they typically require (near-)optimal policies to be guaranteed, and they require agents to articulate their preferences exactly, in order for the mechanism to be non-manipulable. The first condition poses restrictions on the type of planning methods than can be used; which is particularly important in highly complex sequential decision problems. The second condition poses restrictions on the way the utility functions can be accessed. We discuss the implications of this in Section \ref{sec:solutions}.

\subsection{Individual Rewards}
\label{sec:ir}

Up until now, we have considered situations in which all agents have the same  vector input to the utlity function, ${\bf V}$ under SER and ${\bm \rho}$ under ESR, but may have separate individual utilities, $u_i({\bf V})$ or $u_i({\bm \rho})$, with respect to this vector. We now consider situations in which the rewards, and therefore (expected) return vectors, are different for the individual agents. 

First, we note that we consider only two settings for individual rewards: individual utilities and social choice. This is because when individual rewards are received, even if the utility functions for all agents are the same, the resulting utilities are still individual, and the interest of the agents may still be opposed. 

We observe that individual reward settings may seem similar to the team reward but individual utilities settings, regarding the fact that ultimately the joint policies will be selected on the utilities of the individual agents. However, whether the value (or return) vectors are identical or not, can have a profound impact on how complex it is to solve the decision problem. Specifically, a joint policy can often be excluded from consideration if all agents agree that executing a different policy would be better for all agents.\footnote{Note that this is not a sufficient condition in multi-agent settings though, as there may be equilibria that are Pareto-dominated.} When the rewards are shared, all agents will share the same joint policy outcomes and will thus always agree on whether a policy is Pareto-dominated or not. When the rewards are individual however, a joint policy can be the only Pareto-optimal policy (in terms of value or return vectors) for one agent, while it is dominated for another. In other words, settings with individual rewards are considerably more difficult to solve.

\subsubsection{Individual Reward Individual Utility}
\label{sec:iriu}
\begin{figure}[ht]
\centering
\includegraphics[width=0.6\textwidth]{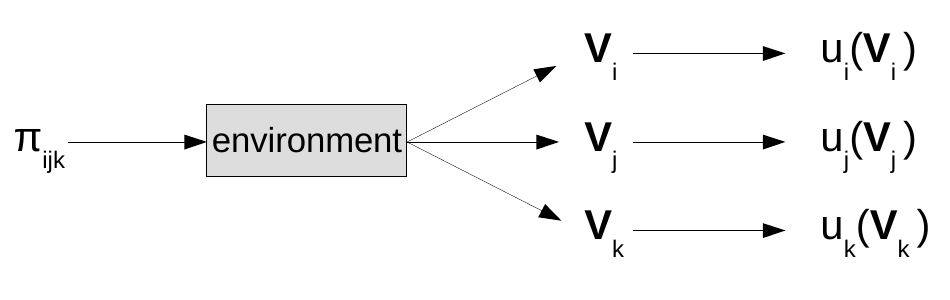}
\caption{The execution phase for the Individual Reward Individual Utility setting. This figure depicts the SER optimality criterion, where the expected values for each agent (i.e., the average over many executions of the policies) will be input to $u$. Under ESR the input to $u$ would be ${\bm \rho}_i$, i.e., the returns for an individual roll-out for an individual agent.}
\label{fig:iriu}
\end{figure}

First let us consider the completely self-interested setting of agents with individual rewards and utilities. This results in the execution phase depicted in Figure \ref{fig:iriu}.

For example \cite{fernandez2002core} study cooperative games, in which coalitions of agents are formed that can obtain rewards in different objectives, and then divide the value of these objectives amongst themselves, leading to individual rewards. Subsequently, they consider what information regarding the utility functions of the agents is available, and whether stable coalitions can be found given this information. 

Because the individual rewards and individual utilities setting is highly complex, it is vitally important to exploit all available information there is with regards to the utility functions of the agents. 
For example, consider the situation in which all individual agents have the same utility function \cite{fernandez2002core,tanino2012vector}, but it is not a priori clear what this utility function is, or the utility function is not fixed. For example, consider the case in which the objectives correspond to resources that can be sold on an open market. Because these prices can vary (possibly rapidly) over time, the agents will need to adjust their policies according to the latest possible price information. A multi-objective multi-agent model with individual rewards and individual utilities, may then help to predict how the agents will respond to changing prices. 

In general, the individual utility functions may be different for each agent, and various degrees of knowledge may exist about their shape or properties. In such settings, it may be hard to produce a sufficiently compact set of possibly viable joint policies to choose from or negotiate with. In this case, we suspect that interactive approaches \cite{igarashi2017multi}, in which more information about the utility functions is actively pursued by querying the agents while planning or learning to limit the set of viable alternatives, will play an important role in future research.  

\subsubsection{Individual Reward and Social Choice with Respect to Individual Utilities}
\label{sec:sw2}

\begin{figure}[ht]
\centering
\includegraphics[width=1\textwidth]{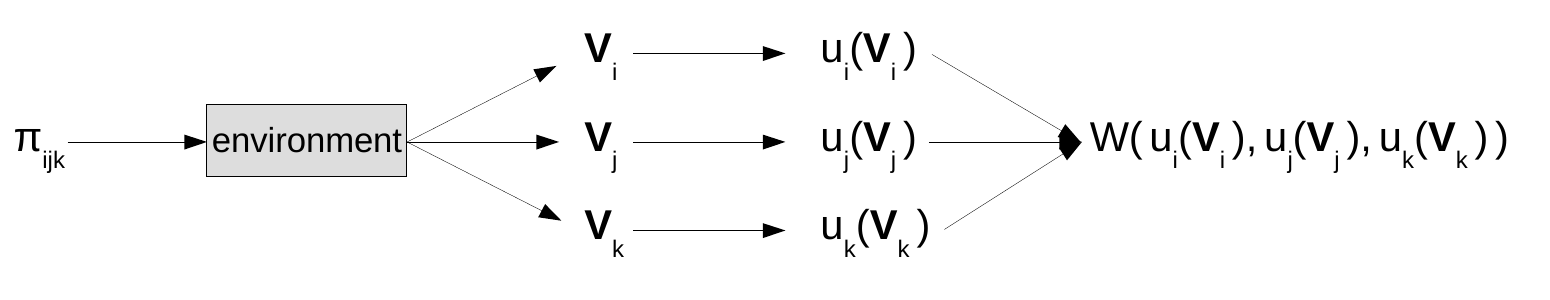}
\caption{The execution phase for the Individual Reward and Social Welfare with Respect to Individual Utilities setting. Please note that the social welfare can depend \emph{both} on the utilities of the agents \emph{and} the value/return vectors for each agent. This figure depicts the SER optimality criterion, where the expected values for each agent (i.e., the average over many executions of the policies) will be input to $u$. Under ESR the input to $u$ would be ${\bm \rho}_i$, i.e., the returns for an individual roll-out for an individual agent.}
\label{fig:iriusw}
\end{figure}

Finally, let us consider the individual rewards and utilities, from the perspective of social choice. This leads to the situation in Figure \ref{fig:iriusw}, in which agents obtain individual value or return vectors, value these according to individual utilities, which are then weighed up, together with the individual value or return vectors, through a social welfare function.

As in the team reward setting, it is important to note that the social welfare function can depend both on the individual value or return vectors, as well as the individual utilities of the agents. For example, in auctions \cite{pla2012multi}, social welfare may depend on attributes of the winning bid(s), as well as a fair outcome in terms of payments to the individual agents, that together the costs the agents need to make to execute their bids if chosen, typically determine the individual utilities.

As in the team reward but individual utilities case, we aim to find a mechanism, i.e., a social welfare function, that forces agents to be truthful about their utility functions, such that the joint policy can be optimised with respect to a notion of social welfare. An interesting -- but to our knowledge unexplored -- aspect would be to investigate, in the case when individual reward vectors are common knowledge, but the preferences are (partially) unknown, whether such mechanisms could still be established, possibly through active querying to obtain information about the individual utility functions. 

\section{Solution Concepts}
\label{sec:solutions}
\begin{figure}
\begin{tikzpicture}
      \matrix (mat) [table] { &  &  \\
         &  &  \\
         &  &  \\
        |[left color=white, right color=white]| &  &  \\
        |[left color=white, right color=white]| &  &  \\
        |[left color=white, right color=white]| &  &  \\
    };
    
\node at ([yshift=0pt]mat-2-1) {\cbox{Coverage sets}};

\node at ([yshift=0pt]mat-2-2) {\cbox{Mechanism design}};
  
\node at ([yshift=-5pt]mat-1-3) {\cbox{Coverage sets  \\ (+ Negotiation)}};
    \node at ([yshift=5pt]mat-3-3) {\cbox{Equilibria and stability concepts}};

  \node at ([yshift=-5pt]mat-4-1) {\cbox{}};

\node at ([yshift=0pt]mat-5-2) {\cbox{Mechanism design}};

\node at ([yshift=-5pt]mat-4-3) {\cbox{Equilibria and stability concepts}};
  
    \node at ([yshift=5pt]mat-6-3) {\cbox{Coverage Sets as best responses}};

\draw[white] (mat-4-1.north west) -- (mat-4-3.north east);
\foreach \col in {1,2,3}
   \draw[white] (mat-1-\col.north west) -- (mat-6-\col.south west);

\node[rotate = 90] at ([yshift=10pt, xshift=-60pt]mat-3-1.north)
    {\textsc{Team}};
    
      \node[rotate = 90] at ([yshift=10pt, xshift=-60pt]mat-6-1.north)
    {\textsc{Individual}};
    
      \node[rotate = 90] at ([xshift=-80pt]mat-4-1.north)
    {\textsc{\large Reward}};
    
  \node at ([yshift=10pt, xshift=0pt]mat-1-1.north)
    {\textsc{Team}};
    
      \node at ([yshift=10pt, xshift=2pt]mat-1-2.north)
    {\textsc{Social choice}};
    
      \node at ([yshift=10pt, xshift=2pt]mat-1-3.north)
    {\textsc{Individual}};

  \node at ([yshift=30pt, xshift=0pt]mat-1-2.north)
    {\textsc{\large Utility}};
\end{tikzpicture}
\caption{Taxonomy mapping to solution concepts for multi-objective multi-agent system}
\label{fig:taxonomy_solconcepts}
\end{figure}

In this section we introduce some of the main solution concepts which are featured in MAS and multi-objective optimisation research, as well as explaining how they relate to the scenarios described in our taxonomy above.

In the context of MAS, it is difficult to identify what constitutes an optimal behaviour, as the agents' strategies are interrelated, each decision depending on the choices of the others. For this reason, we usually try to determine interesting groups of outcomes (i.e., solution concepts), to determine when the system can reach some form of equilibrium. 
Fig. \ref{fig:taxonomy_solconcepts} provides an overview of which of these solution concepts are relevant to each of the five settings in our multi-agent decision making taxonomy.

\subsection{Policies}
\label{sec:policies}
We introduce a few preliminary definitions regarding types of behaviour agents can learn, depending on the action selection strategy given a certain state or on whether or not time plays any role in the policy definition. 

A \textbf{deterministic} (or \textbf{pure}) policy is one where the same action $a$ is always selected for a given state $s$ (i.e., $Pr(a|s)=1$).
A \textbf{stochastic} policy is one where actions in a given state are selected according to a probability distribution (i.e., $Pr(a|s) \in [0,1], \forall a \in A$).
The output of a \textbf{stationary} policy depends only on state, not on time.
The output of a \textbf{non-stationary} policy may vary with both state and time. 

A deterministic game is one where the transition function is deterministic, i.e., the game always transitions to the same next state, for a given system state and joint action. While in single-objective decision problems it is often sufficient to take only deterministic stationary policies into account, it is known that in multi-objective decision problems stochastic or non-stationary policies can lead to better utility \cite{roijers2013survey,white1982multi} both under SER and ESR \cite{roijers2018multi}.

A \textbf{mixture} policy \cite{Shelton2001PhD,vamplew2009constructing} is a stochastic combination of deterministic policies (referred to as base policies). This technique has been used in single agent multi-objective settings to combine two or more deterministic Pareto optimal policies to satisfy a user's preferences. Mixing happens \textit{inter-episode} only, rather than \textit{intra-episode}. Vamplew et al. \cite{vamplew2009constructing} note that switching between base policies during an episode will likely result in erratic and sub-optimal behaviour. Therefore, one of the available base policies is selected probabilistically at the beginning of each episode and followed for the entire episode duration. The aim is to determine mixture probabilities, which on average, after a large number of runs, will yield the desired long-term average return on each objective.

As noted in Section \ref{sec:trtu}, the team reward team utility setting is similar enough to single-agent multi-objective settings such that methods developed for one may be easily applied to the other; mixture policies are one such method which could feasibly be used in the \textbf{team reward team utility setting}. 

\subsection{Coverage Sets}
\label{sec:cs}

The optimal solution in single-agent multi-objective decision making is called a \emph{coverage set (CS)} \cite{roijers2013survey,roijers2017multi}. A coverage set contains at least one optimal policy for each possible utility function, $u({\bf V}^\pi)$, i.e., if a set $\mathcal{C}$ is a coverage set then, under SER,
\[
\forall u\in\mathcal{U} : \max_{\pi \in \Pi} u({\bf V}^\pi) = \max_{\pi \in \mathcal{C}} u({\bf V}^\pi),
\]
and under ESR, 
\[
\forall u\in\mathcal{U} : \max_{\pi \in \Pi} \mathbb{E}[u(\bm{\rho})|\pi, \mu_0] = \max_{\pi \in \mathcal{C}} \mathbb{E}[u(\bm{\rho})|\pi, \mu_0],
\]
where $\Pi$ is the space of all possible (and allowed) policies, $\bm{\rho}$ are the vector-valued returns, i.e., $\bm{\rho} = \sum_{t=0}^\infty \gamma^t {\bf r_t} $ and $\mathcal{U}$ is the set of all possible utility functions. Furthermore, coverage sets do not contain dominated policies, 
\[
\pi \in \mathcal{C} \rightarrow \exists u\in\mathcal{U} : u({\bf V}^\pi) = \max_{\pi' \in \mathcal{C}} u({\bf V}^{\pi'}),
\]
under SER, and, 
\[
\pi \in \mathcal{C} \rightarrow \exists u\in\mathcal{U} : \mathbb{E}[u(\bm{\rho})|\pi, \mu_0] = \max_{\pi' \in \mathcal{C}}\mathbb{E}[u(\bm{\rho})|\pi', \mu_0],
\]
under ESR, i.e., a coverage set should only contain policies that are optimal for some utility function $u$. Finally, algorithms should aim to construct coverage sets that are as small as possible, but as  coverage sets are not unique, so constructing a minimally sized one is far from trivial.

\subsubsection{Motivations for Coverage Sets in Multi-Agent Settings}

In single-agent settings, coverage sets need to be constructed with respect to any possible utility function allowed by the problem specification. However, due to the single-agent nature, it can be assumed that ultimately, in the execution phase, there will be one true utility function that governs user utility. Multi-agent settings are more complex; the different agents can represent different interests, and may be optimising for different utility functions. Nonetheless, there are many multi-agent settings for which coverage sets are the appropriate solution concept. 

The first and most straightforward motivation is the \textbf{team reward and team utility} setting described in Section \ref{sec:trtu}. This is a fully cooperative setting; all rewards and the utility derived from that is shared between all agents. Therefore, there is only one true utility function in the execution phase, and the motivation for coverage sets being the right solution concept is the same as for single-agent multi-objective decision making. For example, this is the case when there are multiple agents belonging to the same team or organisation are tackling a problem together, e.g., a soccer team or different agents belonging to the same company. 

However, team utility is not strictly necessary for coverage sets to be useful. In a team reward but individual utility setting, coverage sets could be used if all agents will agree (preferably contractually) that they will always execute a policy that is potentially optimal. In this case, a coverage set can be computed as the \textbf{input to a negotiation} \cite{jennings2001automated,jonker2017automated} between the agents of which policy to execute. 

Note that this strategy of computing a coverage set and then negotiating does not trivially apply to individual reward settings. In the case of individual rewards, a joint policy can be optimal for one agent, while it can be strictly dominated for another. Generalising the concept of a coverage set to individual reward settings is an open question that would merit investigation. 

Furthermore, in an individual utility setting, a coverage set can also be a \textbf{set of possible best responses to the behaviours of the other agents}. Of course, one needs a different coverage set per combination of possible behaviours for all the individual other agents. This may quickly become infeasible if the set of possible policies of the other agents becomes large. However, if one can model the opponents using a small set of possible behaviours this may be a viable approach. 

Finally, there is a uniquely individual rewards setting coverage set, for the special case that each objective corresponds to one agent, and the objectives of other agents are seen as secondary objectives. In other words, this is the case where agents are at least partially altruistic. This concept is called a Lorenz optimal set, which we discuss in Section \ref{sec:lorenz}.
 
\subsubsection{Convex Coverage Sets}
\label{sec:ccs}

A convex coverage set is the optimal solution set when it can be assumed that the utility functions of all agents are linear. This is a salient case in the multi-objective decision making literature, and for example applies in the case where each objective corresponds to a resource that can be bought or sold on an open market. 
Specifically the utility functions are assumed to be the inner product between a vector of weights $\bf w$ and the value vector of the joint policy ${\bf V}^\pi$, i.e., 
\begin{equation}
u_i({\bf V}^\pi) = {\bf w}_i \cdot {\bf V}^\pi.
\end{equation}
Please note that for this type of utility function, there is no difference between SER and ESR, as $E[{\bf w} \cdot \bm{\rho}|\pi, \mu_0] = {\bf w} \cdot E[\bm{\rho}|\pi, \mu_0] = {\bf w} \cdot {\bf V}^\pi$.

In the case of linear utility functions, the undominated set -- the convex hull -- is defined as follows:
\begin{Def}\label{def:ch}	
	The \emph{convex hull (CH)} is the subset of the set of all admissible joint policies $\Pi$ for which there exists a $\bf w$ for which the linearly scalarised value is maximal:
	\begin{equation}\label{eq:ch}
	CH(\Pi) = \{\pi : \pi \in \Pi \wedge \exists {\bf w} \forall{(\pi' \in \Pi)}~ {\bf w}\cdot {\bf V}^\pi \geq {\bf w}\cdot {\bf V}^{\pi'}\}.
	\end{equation}
\end{Def}
Please note that in this definition, we assume a team reward setting, such that ${\bf V}^\pi$ is a single vector. 

One problem with the CH is that it can be undesirably large; and in the case of stochastic policies, often infinitely large. However, in such cases a \emph{convex coverage set (CCS)} can often been defined that is much more compact\footnote{For details on why this is so, please refer to \cite{roijers2013survey}.}:
\begin{Def}\label{def:ccs}
	A set $CCS(\Pi)$ is a \emph{convex coverage set (CCS)} if it is a subset of $CH(\Pi)$ and if, for every {\bf w}, it contains a policy whose linearly scalarised value is maximal:
	\begin{equation}\label{eq:ccs}
	CCS(\Pi)\subseteq CH(\Pi) \wedge (\forall {\bf w}) (\exists \pi)\left( \pi \in CCS(\Pi) \wedge \forall{(\pi' \in \Pi)}~ {\bf w}\cdot {\bf V}^\pi \geq {\bf w}\cdot {\bf V}^{\pi'}\right).
	\end{equation}
\end{Def}

While in the case of individual utility, the actual ${\bf w}_i$ can differ per agent, the CCS contains at least one optimal policy for every ${\bf w}_i$, and therefore forms a suitable starting point for finding possible compromises based on the assumption that all agents have a linear utility function. For example, a strategy could be to try to estimate each ${\bf w}_i$ and take the average weights vector across all agents to select the default compromise. Of course, agents may also want to negotiate \cite{jennings2001automated,jonker2017automated} in order to get a better deal than such a default compromise.

\subsubsection{Pareto Coverage Sets}

For monotonically increasing but non-linear utility functions, the undominated and coverage sets become significantly larger than for linear utility functions. To be able to define a coverage set for this setting under SER we must first define the concept of Pareto dominance:
\begin{Def}\label{def:pdom}
	A joint policy $\pi$ \emph{Pareto-dominates} another joint policy $\pi'$ when its value is at least as high in all objectives and strictly higher in at least one objective:
	\begin{equation} \label{eq:pdom}
	{\bf V}^\pi \succ_P {\bf V}^{\pi'} \Leftrightarrow \forall i, V^\pi_i \geq V^{\pi'}_i \wedge \exists i, V^\pi_i > V^{\pi'}_i.
	\end{equation}
\end{Def}
Looking at this definition, it is clear that no Pareto-dominated policy can ever have a higher utility under a monotonically increasing utility function:
\[
	{\bf V}^\pi \succ_P {\bf V}^{\pi'} \rightarrow u_i({\bf V}^\pi) \geq u_i({\bf V}^{\pi'}).
\]
As long as being monotonically increasing is the only assumption we can make about the utility function, this is in fact the only thing that can be said of the relative preferences across all possible utility functions. 
Therefore, we use the concept of Pareto dominance to define the undominated set for monotonically increasing utility functions, the Pareto front:
\begin{Def}\label{def:PF}
	The \emph{Pareto front} is the set of all joint policies that are not Pareto dominated by any other joint policy in the set of all admissible joint policies $\Pi$:
	\begin{equation}\label{eq:pf}
	PF(\Pi) = \{\pi : \pi\in\Pi \wedge \neg\exists (\pi'\in\Pi), {\bf V}^{\pi'} \succ_P {\bf V}^\pi\}.
	\end{equation}
\end{Def}
A \emph{Pareto coverage set} (PCS) of minimal size can be constructed by selecting only one policy of the policies with identical value vectors from the $PF(\Pi)$:
\begin{Def}\label{def:pcs}
	A set $PCS(\Pi)$ is a \emph{Pareto coverage set} if it is a subset of $PF(\Pi)$ and if, for every policy $\pi'\in \Pi$, it contains at least one  policy that either dominates $\pi'$ or has equal value to $\pi'$:
	\begin{equation}\label{eq:pcs}
	PCS(\Pi)\!\subseteq\! PF(\Pi) \wedge \forall{(\pi'\! \in\! \Pi)} (\exists \pi)\!\left( \pi\! \in\! PCS(\Pi) \wedge ({\bf V}^\pi\! \succ_P\!  {\bf V}^{\pi'} \vee {\bf V}^\pi\!=\! {\bf V}^{\pi'} )\right)\!.
	\end{equation}
\end{Def}
Negotiating a good compromise from a set of alternatives with different values for all objectives is a typical setting for negotiation \cite{jonker2017automated}. Note that in multi-objective settings, agents and/or users are often incapable of specifying their utilities numerically \cite{zintgraf2018ordered}. However, recently there has been research in automated negotiation focusing on preference uncertainty \cite{tsimpoukis2018automated}, i.e., uncertainty about the individual utility functions, and eliciting preferences \cite{baarslag2017value}, making realistic negotiation with the PCS of a multi-objective decision problem as input, possible. 

Under ESR the situation becomes significantly more complex, i.e., in general, the undominated set is defined as:
\begin{Def}\label{def:unlinU}	
	The \emph{undominated set of policies (U)} under possibly non-linear monotonically increasing $u$, under ESR, is the subset of the set of all admissible joint policies $\Pi$ for which there exists a ${\bf u}$ for which the scalarised value is maximal:
	\begin{equation}\label{eq:unlinU}
	U(\Pi) = \{\pi : \pi \in \Pi \wedge \exists ({\bf u}\!\in\!\mathcal{U})~ \forall{(\pi'\!\in\!\Pi)}~ \mathbb{E}[u(\bm{\rho})|\pi, \mu_0] \geq \mathbb{E}[u(\bm{\rho})|\pi', \mu_0]\}.
	\end{equation}
\end{Def}
However, this is very hard to determine without further information about $u$. To our knowledge no research has yet been done into constructing (approximate) undominated or coverage sets under ESR. In fact, the available research in single-agent MORL typically assumes that $u$ is known \cite{roijers2018multi}.

\subsubsection{Lorenz Optimal Sets}
\label{sec:lorenz}

A uniquely multi-agent coverage set is the Lorenz optimal set \cite{perny2013approximation}. Underlying this solution concept is the assumption that each objective corresponds to the interest of each individual agent. Furthermore, it is assumed that the interests of the other agents are an objective for every agent. In other words, the agents are at least in part altruistic. Finally, it is assumed that ``more equal'' solutions - we will define this exactly below -- are better if sum of utilities does not increase. This final assumption corresponds to a (rather minimal) concept of fairness. 

The use-case for Lorenz optimal sets is: all agents agree that fair solutions are better, hence a set of possibly fair solutions will be computed, after which the agents will negotiate which solution from this set to select. It is thus vital that the agents can rely on the selected solution being followed and that no individual agent will enrich itself to the detriment of the group. This can either be enforced contractually, or simply by the notion that the group of agents will have to rely on each other in the future, and that agents that do not follow the convention will no longer be allowed to participate in other decision problems in which the same agents will need to cooperate. 

The underlying idea of the Lorenz notion of fairness, is the so-called Robin-Hood transfer: if in a vector objective $i$ has a higher value, $v_i$ than objective $j$, $v_j$, then transferring part of the difference, i.e., setting the value of $v_i$ to $v_i-\delta$ and $v_j$ to $v_j+\delta$, for $0<\delta\leq v_i-v_j$, yields a fairer, and therefore better value vector. More formally, this can be captured in the concept of Lorenz domination. To test whether a vector ${\bf V}^\pi$ Lorenz dominates a vector ${\bf V}^{\pi'}$, both vectors are first projected to their corresponding Lorenz vectors:
\begin{Def}
\label{def:lorenzvec}
The Lorenz vector ${\bf L}({\bf V}^\pi)$ of a vector ${\bf V}^\pi$ is defined as:
\[
\left( v_{(1)}, ~ v_{(1)} + v_{(2)}, ~ ... ~ ,~ \sum_{i=0}^N v_{(i)} \right), 
\]
where, $v_{(1)} \leq v_{(2)} \leq ... \leq v_{(N)}$, correspond to the values in the vector ${\bf V}^\pi$ sorted in increasing order.
\end{Def}
NB: this definition is under SER. To our knowledge no research has been done with regards to Lorenz optimality under ESR. 

\begin{Def}
\label{def:lorenzdom}
A vector ${\bf V}^\pi$ Lorenz dominates ($\succ_L$) a vector ${\bf V}^{\pi'}$ when:
\[
{\bf V}^\pi \succ_L {\bf V}^{\pi'} \Leftrightarrow {\bf L}({\bf V}^\pi) \succ_P {\bf L}({\bf V}^{\pi'}),
\]
i.e., when the Lorenz vector of  ${\bf V}^\pi$ Pareto dominates the Lorenz vector of  ${\bf V}^{\pi'}$.
\end{Def}

\begin{Def}\label{def:LOS}
	The \emph{Lorenz Optimal Set} is the set of all joint policies that are not Lorenz dominated by any other joint policy in the set of all admissible joint policies $\Pi$:
	\begin{equation}\label{eq:pf}
	LOS(\Pi) = \{\pi : \pi\in\Pi \wedge \neg\exists (\pi'\in\Pi), {\bf V}^{\pi'} \succ_L {\bf V}^\pi\}.
	\end{equation}
\end{Def}
A \emph{Lorenz coverage set} (LCS) of minimal size can be constructed by selecting only one policy of the policies with identical value vectors  from the $LOS(\Pi)$, similar to constructing a PCS from a PF. 
 
\subsection{Nash Equilibria}
When multiple self-interested agents learn and act together in the same environment, it is generally not possible for all agents to receive the maximum possible reward. Therefore, MAS are often designed to converge to a Nash equilibrium \cite{Shoham2007If} (NE). This notion of equilibrium was first introduced by Nash \cite{Nash1951Non}, and is one of the most important concepts used to analyse MAS \cite{Wooldridge01}.

Consider a multi-agent system with $n$ agents, where $\pi = (\pi_1, \ldots, \pi_i, \ldots, \pi_n)$ is their joint policy, with $\pi_i$ representing the stochastic policy of agent $i$. We also define $\pi_{-i} = (\pi_1,\ldots,\pi_{i-1},\pi_{i+1},\ldots,\pi_n)$ to be a joint policy without the policy of agent $i$. We can thus write $\pi=(\pi_{i}, \pi_{-i})$.

 \begin{Def}\label{def:ne}
 A joint policy $\pi^{NE}$ leads to a Nash equilibrium if for each agent $i \in \{1,...,n\}$ and for any alternative policy $\pi_i$:
\begin{equation}
V_i^{(\pi_i^{NE}, \pi_{-i}^{NE})} \geq  V_i^{(\pi_i, \pi_{-i}^{NE})}
\label{eqn:Nash_Equilibrium}
\end{equation}
\end{Def}
Whenever the above inequality holds true for all possible policies and for all agents in a MAS, a Nash equilibrium exists.
In other words, a Nash equilibrium occurs whenever any individual agent cannot improve its own return by changing its behaviour, assuming that all other agents in the MAS continue to behave in the same way.

In cooperative MAS (i.e., the \textbf{team reward} scenario), coordinating agents' actions to achieve the highest possible system welfare is already a difficult problem. While it is possible for multiple individual learners in a cooperative MAS to converge to a point of equilibrium, whether they will converge to an optimal joint policy (one which maximises the system welfare) depends on the specific learning algorithm and reward scheme used.

\subsubsection{Nash Equilibrium in Multi-Objective Multi-agent Settings}

In multi-objective decision making, each agent is trying to optimise his return along a set of objectives. Each agent needs to also make compromises between competing objectives on the basis of his/her utility function. As a motivation for why NE is an appropriate solution concept in MOMAS, we look at the \textbf{team reward individual utility} (section~\ref{sec:triu}) and \textbf{individual reward individual utility} (section \ref{sec:iriu}) scenarios. In both these cases, the utility derived by each agent from the received reward is different, regardless if this reward is the same or not for all the agents. These constitute the most difficult scenarios in our taxonomy. Furthermore, one should also consider which optimisation criteria are best to use, based on what each agent is looking to optimise. 
Depending on whether an agent cares about average performance over a number of policy executions, or just the performance of a single policy execution \cite{roijers2018multi}, we can define the concept of Nash Equilibrium from the perspective of the two multi-objective optimisation criteria defined in Section~\ref{sec:optimisation}: ESR and SER. 

To simplify our notation let us denote the discounted sum of rewards received by agent $i$ by: $\bm{\rho}_i =  \sum\limits^\infty_{t=0} \gamma^t \mathbf{r}_{i,t}$. We can then re-write the expected value of a joint policy $\pi$, given the distribution $\mu_0$ over initial states as: $\bm{V}^{\pi} = \mathbb{E} \left[ \bm{\rho}_i \:|\: \pi, \mu_0 \right]$. 

\begin{Def}[Nash equilibrium for Expected Scalarised Returns]
\label{def:ne_esr}
A joint policy $\pi^{NE}$ leads to a Nash equilibrium under the Expected Scalarised Returns criterion if for each agent $i \in \{1,...,n\}$ and for any alternative policy $\pi_i$:
\begin{equation}
\E  \left[u_i ( \bm{\rho}_i) \:|\: (\pi_i^{NE}, \pi_{-i}^{NE}), \mu_0 \right] \geq \E \left[ u_i (\bm{\rho}_i) \:|\: (\pi_i, \pi_{-i}^{NE}), \mu_0\right]
\label{eqn:ne_esr}
\end{equation}
\noindent i.e., $\pi^{NE}$ is a Nash equilibrium under ESR if no agent can increase the \emph{expected utility of her returns} by deviating unilaterally from $\pi^{NE}$.
\end{Def}

\begin{Def}[Nash equilibrium for Scalarised Expected Returns]
\label{def:ne_ser}
A joint policy $\pi^{NE}$ leads to a Nash equilibrium under SER if for each agent $i \in \{1,...,n\}$ and for any alternative policy $\pi_i$:
\begin{equation}
u_i \left(\mathbb{E} \left[ \bm{\rho}_i \:|\: (\pi_i^{NE}, \pi_{-i}^{NE}), \mu_0 \right] \right) \geq  u_i \left(\mathbb{E} \left[ \bm{\rho}_i \:|\: (\pi_i, \pi_{-i}^{NE}), \mu_0 \right]  \right)
\label{eqn:ne_ser}
\end{equation}
\noindent i.e. $\pi^{NE}$ is a Nash equilibrium under SER if no agent can increase the \emph{utility of her expected returns} by deviating unilaterally from $\pi^{NE}$.
\end{Def}

Under non-linear utility functions, it has been shown that the choice of optimisation criterion can alter the set of Nash equilibria \cite{radulescu2019equilibria}. Furthermore, under SER, even in a multi-objective normal form game, NE need not exist.

\subsection{$\epsilon$-approximate Nash Equilibria}
An $\epsilon$-approximate Nash equilibrium \cite{nisan2007algorithmic} occurs when an individual agent cannot increase its return by more than an additive $\epsilon > 0$ by deviating from its policy, assuming that all other agents continue to behave in the same way. In other words, an agent will not care to switch his policy, if the obtained gain is too small. 

 \begin{Def}\label{def:epsilon_ne}
 A joint policy $\pi^{NE}$ leads to a $\epsilon$-Nash equilibrium if for each agent $i \in \{1,...,n\}$ and for any alternative policy $\pi_i$:
\begin{equation}
V_i^{(\pi_i^{NE}, \pi_{-i}^{NE})} \geq  V_i^{(\pi_i, \pi_{-i}^{NE})} - \epsilon
\label{eqn:epsilon_NE}
\end{equation}
\end{Def}

$\epsilon$-Nash equilibria can be envisioned as regions surrounding any Nash equilibrium. All the definitions for NE under SER and ESR can also be adapted for the case of $\epsilon$-Nash equilibria by subtracting $\epsilon$ from the right side of each inequality. 

\subsection{Correlated Equilibria}
A correlated equilibrium (CE) is a game theoretic solution concept proposed by Aumann \cite{aumann1974subjectivity} in order to capture correlation options available to the agents when some form of communication can be established prior to the action selection phase. Another way to think about this concept is to envision an external device sampling from a given distribution and providing each agent with a private signal (e.g., a recommended action) at each time-step. Given this private signal, each agent can then independently make a decision on how to act next.
For this work we will consider that the signals take the form of action recommendations.

While previously discussed policies define state-based action probabilities independently for each agent, a \emph{correlated policy} $\sigma$ represents a probability distribution over the joint-action space $\mathcal{A}$ of all the agents in the system (i.e., $Pr(\mathbf{a}|s) \in [0,1], \forall \mathbf{a} \in \mathcal{A}$). Thus, correlated policies introduce explicit dependencies between the agents' behaviours. Let us define $\pi^{\sigma}=(\pi^{\sigma}_1, \ldots, \pi^{\sigma}_n)$ as the joint policy of the agents when following the recommendation provided according to a correlated policy $\sigma$.

 \begin{Def}\label{def:ce}	
 A correlated policy $\sigma^{CE}$ is a correlated equilibrium if for each agent $i \in \{1,...,n\}$ with its corresponding policy under $\sigma^{CE}$, $\pi^{\sigma^{CE}}_i$, and for any alternative policy $\pi_i$:
\begin{equation}
V_i^{(\pi^{\sigma^{CE}}_i, \pi^{\sigma^{CE}}_{-i})} \geq  V_i^{(\pi_i, \pi^{\sigma^{CE}}_{-i})}
\label{eqn:Correlated_Equilibrium}
\end{equation}
\end{Def}

Thus, a correlated equilibrium ensures that no player can gain additional return by deviating from the suggestions, given that the other players follow them as well.

\subsubsection{Correlated Equilibria in Multi-Objective Multi-agent Settings} 

Similarly to the Nash equilibria case, solution concepts such as correlated equilibria can be used in scenarios in which each agent derives a different utility from the received reward vector, i.e., the \textbf{team reward individual utility} (section~\ref{sec:triu}) and \textbf{individual reward individual utility} (section \ref{sec:iriu}) settings. Furthermore, we can also define correlated equilibria from the perspective of the two possible optimisation criteria: ESR and SER, when considering multi-objective multi-agent decision making problems. We will again denote the value of a joint policy $\pi$ as $\bm{V}^{\pi} = \mathbb{E} \left[ \bm{\rho}_i \:|\: \pi, \mu_0 \right]$, where $\bm{\rho}_i =  \sum\limits^\infty_{t=0} \gamma^t \mathbf{r}_{i,t}$.

\begin{Def}[Correlated equilibrium for Expected Scalarised Returns]
\label{def:ce_esr}
A correlated policy $\sigma^{CE}$ is a correlated equilibrium under ESR if for any agent $i \in \{1,...,n\}$ with its corresponding policy under $\sigma^{CE}$, $\pi^{\sigma^{CE}}_i$, and for any alternative policy $\pi_i$:
\begin{equation}
\E  \left[u_i ( \bm{\rho}_i) \:|\: (\pi^{\sigma^{CE}}_i, \pi^{\sigma^{CE}}_{-i}), \mu_0 \right] \geq \E \left[ u_i (\bm{\rho}_i) \:|\: (\pi_i, \pi^{\sigma^{CE}}_{-i}), \mu_0\right]
    \label{eqn:ce_esr}
\end{equation}
\noindent i.e. $\sigma^{CE}$ is a correlated equilibrium under ESR if no agent can increase the \emph{expected utility of her returns} by deviating unilaterally from the action recommendations in $\sigma^{CE}$.
\end{Def}

\begin{Def}[Correlated equilibrium for Scalarised Expected Returns]
\label{def:ce_ser}
A correlated policy $\sigma^{CE}$ is a correlated equilibrium under SER if for any agent $i \in \{1,...,n\}$ with its corresponding policy under $\sigma^{CE}$, $\pi^{\sigma^{CE}}_i$, and for any alternative policy $\pi_i$:
\begin{equation}
u_i \left(\mathbb{E} \left[ \bm{\rho}_i \:|\: (\pi^{\sigma^{CE}}_i, \pi^{\sigma^{CE}}_{-i}), \mu_0 \right] \right) \geq  u_i \left(\mathbb{E} \left[ \bm{\rho}_i \:|\: (\pi_i, \pi^{\sigma^{CE}}_{-i}), \mu_0 \right]  \right)
\label{eqn:ce_ser}
\end{equation}
\noindent i.e. $\sigma^{CE}$ is a correlated equilibrium under SER if no agent can increase the \emph{utility of her expected returns} by deviating unilaterally from the given action recommendations in $\sigma^{CE}$. \footnote{When considering a CE-based approach, an agent is able to calculate his expected return given one correlation signal, but also an expected return given all the possible signals. This allows one to define two variants for CE under SER: the single-signal CE (when agents have multiple interactions under the same given signal) and multi-signal CE (when agents receive a new signal after every interaction) \cite{radulescu2019equilibria}. For this work we define the more general case of multi-signal CE.}
\end{Def}

It has also been shown that the set of correlated equilibria will be altered, depending on the optimisation criteria used \cite{radulescu2019equilibria}. Furthermore, under SER, with non-linear utility functions, in multi-objective normal form games, CE need not exist when taking the expectation over all the possible correlation signals.

\subsection{Coalition Formation and Stability Concepts}

A different perspective on multi-agent decisions is that taken by \emph{cooperative game theory} \cite{chalkiadakis2011computational}. Cooperative game theory studies settings where binding agreements among agents are possible. A central problem is therefore that of \emph{coalition formation}, i.e., finding (sub)groups of agents that are willing to make such a binding agreement with each other. In the models in cooperative game theory, the utility for each agent is directly derived from the coalition the agents end up in, however, one can imagine that under the hood, the coalition works together cooperatively (based on their binding agreement) in a sequential decision problem that results in this utility. We further note, that the word cooperative does not imply team utility; typically, the agents will have their own utility functions. Hence, the solution concepts from cooperative game theory apply to the individual utility settings. 

To illustrate the solution concepts for multi-objective cooperative game theory, we use the \emph{multi-criteria coalition formation game (MC2FG)} \cite{igarashi2017multi,tanino2009multiobjective,tanino2012vector}. Such a game consists of a set of agents, $\mathcal{N}$,  each with their own utility function $u_i({\bf q})$, and a quality/reward function ${\bf q}(S)$ that maps each possible subset, i.e., coalition, of the agents $S\in\mathcal{N}$ to a value or quality vector, that each agent in that coalition will receive. That is, we are in an individual utility setting.

\begin{Def} A multi-criteria coalition formation game (MC2FG) is a triple $(N,q,\mathcal{U})$ where $N$ a finite set of agents, ${\bf q} : 2^N\rightarrow\mathbb{R}^d$ is a vector-valued reward function that represents the quality ${\bf q}(S)$ of a subset, i.e.\ coalition, of agents $S\subseteq N$, and ${u}_i \in \mathcal{U}$ are the utility functions for each agent $i\in N$.
\end{Def} 

The MC2FG is a useful model to study for multi-objective multi-agent systems. Specifically, if in a multi-agent system with multiple objectives, the agents need to form coalitions to cooperate to gain a value vector, the most straightforward case is a MC2FG, i.e.,  given the coalition the value vector can exactly be predicted independently of the other coalitions, but agents can have different preferences between possible value vectors. Therefore, MC2FGs form a minimal model to study the feasibility of contract negotiations between agents in multi-objective multi-agent decision making.

The goal in an MC2FG is to find a partition, $\psi$, of agents into coalitions that are stable. That is, the coalitions will not break apart. For this notion of stability, there are multiple possible versions, from strong to weak: \emph{core stability}, \emph{Nash stability}, and \emph{individual stability}. 

We denote the coalition (subset of agents) which agent $i$ is in according to $\psi$ as $\psi(i)$. A partition $\psi$ is \emph{individually rational} if no agent strictly prefers staying alone to their own coalitions, i.e. $\forall i : u_i({\bf q}(\psi(i)) \geq u_i({\bf q}(\{ i \}))$. 

\begin{Def}
A coalition $S\subseteq N$ is said to \emph{block} a partition $\psi$ if every agent strictly prefers $S$ to $\psi(i)$, i.e., $\forall(i\in S) : u_i({\bf q}(\psi(i)) < u_i({\bf q}(S))$.
\end{Def}
\begin{Def}
	A partition $\psi$ of $N$ is \emph{core stable (CR)} if no (non-empty) coalition $S\subseteq N$ blocks $\psi$.
\end{Def}

Beside CR, there are two key stability concepts that represent immunity to deviations by individual players. 
An agent $i$, wants to deviate from $\psi(i)$ to another coalition in $\psi$, $S$, if  it prefers $S\cup\{i\}$ to $\psi(i)$, i.e., $u_i({\bf q}(\psi(i)) < u_i({\bf q}(S\cup\{i\}))$. A player $j\in S$ would accept such a deviation if it prefers $S\cup\{i\}$ to $S$, i.e.,  $u_j({\bf q}(S) \leq u_j({\bf q}(S\cup\{i\}))$. 

\begin{Def} 
	A deviation of $i$ from $\psi(i)$ to $S$ is an NS-deviation if $i$ wants to deviate from $\psi(i)$ to $S$.
\end{Def}
\begin{Def}
	A deviation of $i$ from $\psi(i)$ to $S$ is an IS-deviation if it is an NS-deviation and all players in $S$ accept it.
\end{Def}
\begin{Def} A partition $\psi$ is \emph{Nash stable (NS)} if there are no NS-deviations for any agent $i$, from its coalition $\psi(i)$ to any other coalition $S\in \psi$ or to $\emptyset$. 
\end{Def} 
\begin{Def} A partition $\psi$ is \emph{Individually stable (IS)} if there are no IS-deviations for any agent $i$, from its coalition $\psi(i)$ to any other coalition $S\in \psi$ or to $\emptyset$. 
\end{Def} 

Every single-criterion coalition formation game has at least one partition that is core stable and individually stable \cite{igarashi2017multi}. However, this is not necessarily so in the multi-objective case. This is because in a single-objective coalition formation game, the utility of a coalition is the same for each agent, i.e., the scalar quality/reward of the coalition. However, in the multi-objective case, all agents that are in a coalition $S$ receive the same reward vector ${\bf q}(S)$, but they may value these vectors differently. In fact, Igarashi and Roijers\ \cite{igarashi2017multi} show that MC2FGs do not necessarily have core, Nash, nor individually stable partitions by counter-example resulting in the following Theorem: 
\begin{Theo}
For any positive integer $n$ and for any $0<\varepsilon<1$ there exists an MC2FG $(N,{\bf q},\{{\bf w}_i : i\in N\})$, where ${\bf w}_i$ is the weights vector for the linear utility function of agent $i$, which admits neither a core nor individually stable partition, where the number of players $|N|=n$, the number of criteria $m= 2$, and $| w_{i,k} - w_{i,k}|\leq \varepsilon$ for any $i,j \in N$ and either objective ($k$).
\end{Theo}

So, this theorem implies that even when the number of objectives is smaller than the number of agents, and the difference between the utility functions (even if they are linear) is arbitrarily small, stable partitions do not need to exist. This has important consequences for multi-objective multi-agent systems in general, as MC2FGs are such a minimal model of finding cooperative subsets of agents that could contractually agree on a value vector. Because no stable solutions need to exist, such contract negotiations could go on forever (agents repeatedly switching between coalitions before signing the contract), if all agents just aim to optimise their individual utilities. We believe this means that a thorough investigation of (the compatibility of) negotiation techniques for various multi-objective multi-agent decision problems on the basis of coverage sets, under different optimalisation criteria (i.e., ESR versus SER) is required. Furthermore, the fact that the stability of coalitions cannot be guaranteed could have a strong impact on future \emph{interactive approaches}\footnote{Interactive approaches intertwine preference elicitation and learning about the decision problem \cite{roijers2017interactive,roijers2018interactive}.} as well. While the prospects of such interactive approaches seem good, as \cite{igarashi2017multi} have shown that individually stable coalitions can often be found interactively under linear utility functions in MC2FGs, it is not clear what will happen for non-linear utility functions under SER or ESR, or in learning settings where the estimated value vectors of different joint policies of changing coalitions, may change.  

\subsection{Social Welfare and Mechanism Design}

In this section, we have so far taken the position of the individual agents. However, we can also take a system perspective, i.e., we can look at what the socially desirable outcomes of a multi-agent decision problem would be. In Section \ref{sec:sw1} and \ref{sec:sw2}, we have looked at the execution phase of such settings and defined the social welfare function, i.e., a function that should be maximised if we want to find socially desirable outcomes. 

In game theory, the field of mechanism design takes the system's perspective for multi-agent decision problems: taking an original decision problem where the agents have individual reward functions that are unknown to the other agents and the ``owner'' of the game, as well as a social welfare function as input, the aim is to design a system of additional payments that would a) force the agents to be truthful about their individual utilities, and b) leads to solutions that are (approximately) optimal under the social welfare function. 

In single-objective multi-agent decision problems, the individual utilities of the agents are simply the individual (expected) (cumulative) rewards that the agents receive. The agents can be assumed to know these rewards, and act accordingly. This is for example the case in public tenders, where different companies know their own costs and profit margins of their possible proposals, but do not broadcast this information to others. In multi-objective settings, the situation is more complex, as the individually received rewards determine the individual utilities via individual private utility functions. These utility functions can have different properties. In general, it might even be very hard, or even impossible to articulate these functions, so being ``truthful'' about their utilities might be infeasible from the get-go. 

Nevertheless, it is possible to design mechanisms for some multi-objective multi-agent problems if the individual utilities can be articulated. First, we observe that if the utility functions are linear, the inner product with weights distributes over all expectations. Hence, it is possible to even design mechanisms that are agnostic about the weights, compute a convex coverage set (see Section \ref{sec:ccs}) of possibly socially desirable outcomes, and choose the weights a posteriori. This enables the designer/owner of the decision problem to make an informed decision about which weights to use. For example, in a public tender for traffic maintenance by Scharpff et al.\  \cite{scharpff2013coordinating,Roijers2014Linear}, the objectives of costs and traffic hindrance should both be minimised. Because of mechanism design, all agents need to be truthful; whatever weights (and resulting penalties) are put on traffic hindrance, it is in the best interest of the agents to be truthful about their costs, making it possible for the owner of the game to assume that given the mechanism, all agents will be fully cooperative, solve the problem as an MOMMDP, and choose the weights a posteriori. 

For specific cases of non-linear utility functions, it is also possible to devise mechanisms. For example, Grandoni et al.\ \cite{grandoni2010utilitarian} assume individual utility functions with a primary objective that should be maximised, and other objectives that need to achieve at least a threshold value. The utility is the value of the first objective in the case that all thresholds are met, but negative infinity if the thresholds are not met. They show that for such cases, effective mechanisms can be designed, and solutions can be found within a reasonable amount of time. 

An interesting and different approach to social welfare is taken by Mouaddib et al. \cite{mouaddib2007towards}, who cast a decentralised sequential multi-agent problem with individual (scalar) reward functions as a multi-objective problem. Specifically, besides its main objective an agent will model its positive impact on the group as well as the nuisance it causes to other group members as separate objectives. Even though this work provides no strong guarantees, the authors show empirically that these additional objectives in combination with a social welfare function can lead to good emergent group behaviour in very hard -- decentralised partially observable multi-agent -- decision problems.

\subsection{Other Solution Concepts}
The concepts discussed so far do not form in any way an exhaustive list for what constitutes a solution in a MOMAS. We briefly present below a few other possible solution concepts that have been discussed in the literature.

An early discussion on how to extend equilibria concepts from single-objective games to multi-objective settings can be found in \cite{shapley1959equilibrium}, where the concepts of weak and strong equilibria are proposed as extensions of NE. These concepts are defined using vector domination and thus are called \emph{Pareto-Nash Equilibria} \cite{borm1988pareto,voorneveld1999axiomatizations,lozovanu2005multiobjective}. Continuing the game theoretic perspective, \cite{kawamura2013evolutionarily} extends the concept of \emph{evolutionary stability} for multi-objective games.

\emph{Cyclic equilibria} \cite{flesch1997cyclic,mirrokni2004convergence,zinkevich2006cyclic} have been proposed as a solution concept for games where no stationary equilibrium exists. A cyclic equilibrium is a non-stationary joint policy where agents have no incentive to deviate unilaterally \cite{zinkevich2006cyclic}. Cyclic equilibria cycle repeatedly through a set of stationary policies. Similar to $\epsilon$-NE, an $\epsilon$-correlated cyclic equilibrium is defined as a situation where no agent can improve its value by more than $\epsilon$ at any stage by deviating unilaterally \cite{zinkevich2006cyclic}.
 
\section{Algorithmic Approaches and Applications }
\label{sec:algorithmic}

In the section, we survey related work on algorithmic approaches to MOMAS, as well as applications of multi-objective multi-agent decision making. The works we survey are organised into three broad categories: those which aim to derive coverage sets are discussed in Section \ref{sec:coverage_sets}, those which aim to apply stability and equilibria concepts are discussed in Section \ref{sec:equilibria_stability} and finally methods which employ mechanism design are discussed in Section \ref{sec:mec_design}. Within these categories, the works surveyed are further classified according to our taxonomy on the basis of the reward and utility functions used.

\subsection{Coverage Sets}
\label{sec:coverage_sets}

\subsubsection{Team Reward - Team Utility (TRTU)}

 Multi-objective coordination graphs (MOCoGs)\footnote{Because the MOCoG model is a flexible multi-objective graphical model that can be used for many types of problems, and has been used by many research communities, the MOCoG is known under many different names. These include: multi-objective weighted constraint satisfaction problems (MO-WCSPs) \cite{Rollon06MOBE} and Multi-objective Constraint Optimisation Problems (MOCOPs) \cite{marinescu2009exploiting,marinescu2013multi}.} are one of the most studied models for cooperative multi-objective multi-agent systems, and in particular for team reward team utility. One reason for this is that it is the simplest model to express and exploit \emph{loose couplings}, i.e., the fact that in multi-agent systems the rewards can often be factorised into a sum over small components, i.e., local reward functions, that depend on small (but possibly overlapping) subsets of agents. However, finding suitable joint actions in a MOCoG is also key to finding, e.g., coverage sets for MOMMDPs (as is also the case for single-objective CoGs and MMDPs \cite{KokVlassis2004,kok2006collaborative}). 
 
 In MOCoGs, a lot of research focuses on finding (approximate) Pareto coverage sets (PCSs), using various algorithmic approaches. These approaches often extend  single-objective methods by adapting the inner workings of such methods to be able to handle sets of value vectors rather than single scalar values. Examples of such methods are multi-objective bucket elimination (MOBE) \cite{Rollon06MOBE,rollonThesis}, also known as multi-objective variable elimination (MOVE, which is the more common in the planning and reinforcement learning communities), which  solves a series of local sub-problems to eliminate all agents from a MOCoG in sequence, by finding local coverage sets as best responses to neighbouring agents. Other such methods include multi-objective Russian doll search \cite{rollon2007multi}, multi-objective (AND/OR) branch-and-bound tree search \cite{marinescu2009exploiting,marinescu2011efficient,rollon2008constraint}  using mini-bucket heurtistics \cite{Rollon06MOBE,marinescu2009exploiting},  Pareto local search \cite{inja2014queued}, and multi-objective max-sum \cite{delle2011bounded}. Many of these papers note that PCSs can grow very large very quickly, making finding exact PCSs infeasible. Therefore computing bounded approximations \cite{marinescu2011efficient} can be necessary. 
 
 On the other hand, Roijers et al. \cite{Roijers2013Computing,Roijers2014Linear,roijers2015computing} compare the computational and memory complexity of computing convex and Pareto coverage sets (CCSs and PCSs). They observe that the size of PCSs typically grows much faster with the number of agents in a MOCoG than the size of CCSs, and that often CCSs suffice, e.g., in the case that mixture policies are allowed (Section \ref{sec:policies}). It can therefore be highly preferable to focus on finding CCSs, especially in problems with many agents. They propose several methods to do so with different computation-memory trade-offs. Specifically for finding CCSs, they propose and compare inner loop methods to outer loop methods, i.e., methods that construct CCSs by iteratively solving scalarised instances of the multi-objective decision problem. Outer loop methods are more memory-efficient, and significantly faster for smaller numbers of objectives, while inner loop methods are faster for many objectives. Finally, they note that $\varepsilon$-approximate CCS can efficiently be computed using outer loop methods. Anytime approximations to CCSs can also be effectively constructed using an outer loop method that employs variational (inference) methods \cite{roijers2015variational}. 
 
Wilson et al.\ \cite{wilson2015} consider methods to compute coverage sets for MOCoGs when more information about the possible utility function(s) is available. They assume that along with the standard notion of the shape of the utility function\footnote{Or, in their original paper, the equivalent concept of domination.}, they are provided with a set of preferences that users expressed a priori. They integrate these preferences into AND/OR branch-and-bound, and show that this can often lead to much more efficient computation than would be required to compute a PCS. To achieve this, they pose additional constraints on the utility function, that are only fulfilled by linear utility functions. Therefore, doing the same for arbitrarily shaped utility functions is still an open problem. 

Multi-objective multi-agent MDPs (MOMMDPs) -- in the literature often referred to as cooperative multi-objective stochastic games (cooperative MOSGs) -- are another frequently used model for cooperative multi-objective multi-agent decision making problems. Some recent works have sought to derive coverage sets in MOMMDPs using reinforcement learning or evolutionary algorithms (e.g. \cite{Yliniemi2015PhD,Yliniemi2016Multi,Mannion2016Dynamic,Mannion2017Policy,Mannion2018Reward,Mannion2017PhD}). As in single-objective MMDPs, learning joint policies which coordinate agents' actions to get the desired outcome(s) in MOMMDPs is a difficult problem. When individual agents learn using the system reward and the same utility function (i.e TRTU), it is difficult for any one agent to evaluate how its actions affected the system utility, due to the effect of the other agents in the system. This is referred to as the credit assignment problem in the single-objective MAS literature. Reward shaping is one solution which has been proposed to address this problem, where the reward which is usually received from the environment is augmented with an additional shaping term, with the goal of providing a more informative reward signal to the agents in a MAS. Specific forms of reward shaping which have been applied to cooperative MOMAS include difference rewards ($D$) \cite{Wolpert2000Collective} and potential-based reward shaping ($PBRS$) \cite{Ng99}\footnote{Although individualised reward shaping implies that each agent receives a different reward, we have classified these works under the TRTU setting as all agents use the same utility function and the individual shaped rewards are still aligned with the global (system) rewards. Reward shaping might also be useful in combination some of the other settings in our taxonomy and solution concepts discussed in Section \ref{sec:solutions}, although only the TRTU setting with coverage sets has been explored to date.}. 

Yliniemi \cite{Yliniemi2015PhD} and Yliniemi and Tumer \cite{Yliniemi2016Multi} present the first work that considers the use of reward shaping in a cooperative MOMARL setting. Their work compares the effectiveness of the difference reward with that of two typical MARL reward structures: local rewards ($L$) and global rewards ($G$). Experimental work in a multi-objective congestion problem, and a multi-objective robot coordination problem confirms that $D$ can improve MOMARL performance when compared to $L$ or $G$, both in terms of learning speed and the quality of the set of non-dominated solutions found. \cite{Yliniemi2016Multi} also demonstrates that $D$ can be used effectively with multi-objective evolutionary algorithms, in a series of experiments where it is applied to shape the fitness function of the Elitist Non-dominated Sorting Genetic Algorithm-II (NSGA-II). \cite{Mannion2016Multi} evaluate the effect of $D$ in an electricity generator scheduling problem. 
Mannion et al. \cite{Mannion2017Policy} provide a theoretical analysis of $PBRS$ in single- and multi-agent MORL settings, demonstrating that the Pareto relation between (joint) policies is preserved when $PBRS$ is used. Mannion et al. \cite{Mannion2017Theoretical} also provide a theoretical analysis of $D$ in MOMMDPs, demonstrating that the relative values of actions (and therefore the Pareto relation between actions) is preserved when $D$ is applied in MOMMDPs. A comprehensive analysis of the effects of both $D$ and $PBRS$ when generating coverage sets via learning in cooperative MOMARL settings is presented in \cite{Mannion2018Reward} and \cite{Mannion2017PhD}.

On the application side, Ahmad et al.\ \cite{ahmad2008using} consider the problem of multi-core processing, or, more specifically, energy aware task allocation for optimising energy use versus performance. They employ a cooperative game theory perspective and transform the problem into a max-max-min one to generate solutions for different energy-time trade-offs. Bone and Dragi{\'c}evi{\'c} \cite{bone2009gis} are interested in the setting of natural resource management, where each agent represents a forest harvesting company. Agents need to learn how to harvest wood in order to maximise economic profit and minimise ecological impact. They use a step utility function to transform the problem into a single-objective one and learn an optimal policy as independent reinforcement learners. Babbar-Sebens and Mukhopadhyay \cite{babbar2009reinforcement} consider a water resource management system modelled as a multi-objective game, where the players use a simple reinforcement learning or genetic algorithm approach in order to find a set of solutions corresponding to various linear utility functions. Focusing on traffic optimisation, Houli et al.\ \cite{houli2010multiobjective} develop the multi-RL algorithm, a multi-agent reinforcement learning approach which selects an optimisation objective depending on the real-time traffic conditions.

\subsubsection{Team Reward - Individual Utility (TRIU)} 

Aoki et al.\ \cite{aoki2004distributed} model a multi-stage flow systems as a MAS, where each agent (i.e., service centre) represents a different objective. They use a distributed reinforcement learning framework and propose a bi-directional decision making mechanism to address the multi-objective nature of the problem. 

\subsubsection{Individual Reward - Individual Utility (IRIU)}
Investigating multi-objective games, Avigad et al.\ \cite{avigad2011optimal} propose an evolutionary search algorithm for finding the Pareto set of strategies for a player, given each possible strategy of an opponent. Also looking at competitive multi-objective games, Eisenstadt et al.\ \cite{eisenstadt2015co} proposes a co-evolutionary algorithm for finding solutions simultaneously for both players, under worst-case assumptions, given that the oponent's preferences are unknown.

Brys et al. \cite{Brys2014Distributed} apply MOMARL to a traffic signal control problem, where each intersection in a $2 \times 2$ grid is controlled by an individual agent. Their work demonstrats that rewarding agents with a linear scalarised combination of delay and throughput can improve delay times when compared to agents rewarded using delay alone. However, their approach uses local rewards (i.e. each agent is rewarded based on conditions at its assigned intersection only, and hence this work is classified as IRIU), and does not make any attempt to explicitly encourage coordination between the agents.

Dusparic and Cahill \cite{dusparic2009distributed} propose the Distributed W-Learning algorithm, an RL-based approach for multi-policy optimisation in collaborative multi-agent systems, such as urban traffic. Each agent represents a traffic light at an intersection and implements a set of local and remote policies (i.e., involving its neighbours). Even though the agents here have possibly conflicting goals and receive an individual reward, at every time-step they exchange information with their neighbours regarding their current states and rewards, allowing them to develop, if necessary, a cooperative behaviour.

Van Moffaert et al. \cite{vanMoffaert2014Novel} apply MORL to a multi-objective multi-agent smart camera problem. They develop an adaptive weight algorithm (AWA) which is used to choose the weighting between the two system objectives when linear scalarisation is applied to individual reward vectors for each camera agent in the system. The AWA algorithm is found to improve learning speed, obtaining solutions with a better spread in the objective space, when compared with other weight selection methods that were tested.

\subsection{Equilibria and Stability methods}
\label{sec:equilibria_stability}
\subsubsection{Team Reward - Individual Utility (TRIU)}

Lee \cite{lee2012multi} takes a game theory perspective on the reservoir watershed management domain and develops for this purpose a multi-objective game theory model. The goal in this case is to balance between maximising economic gain and minimising environmental impact. Each player represents a different objective and multiple bargaining rounds are used in order to arrive at a Nash equilibrium.

\subsubsection{Individual Reward - Individual Utility (IRIU)}

Qu et al. \cite{qu2015robust} examine multi-objective non-zero sum game, where each agent has a set of weights for each objectives (i.e., linear utility function where the exact weights are not known). They propose an approach for finding a robust weighted Nash Equilibrium.

Taylor et al.\ \cite{Taylor2014Accelerating} propose Parallel Transfer Learning (PTL) as a mechanism to accelerate learning, by sharing experience among agents. PTL is tested on a multi-objective multi-agent smart grid problem, and is found to improve learning speed and final performance when compared to agents learning without PTL.

Madani and Lund \cite{madani2011monte} consider a Monte Carlo Game Theory approach for stochastic multi-criteria decision making settings, such as water resource management. They propose the use of Monte Carlo simulations to map the problem to deterministic settings transformed then into strategic games, solved using non-cooperative stability methods.

\subsection{Social Welfare and Mechanism Design}
\label{sec:mec_design}

On the line of reward engineering in multi-objective congestion problems, Ramos et al.\ \cite{ramos2019budget} consider the route choice traffic problem and develop a reward signal based on the marginal cost tolling mechanism. This allows one to reach a system optimum performance, even when agents have heterogeneous preferences with respect to travelling time and monetary costs. To model the agent preferences they use a linear utility function that attributes different weights to the considered objectives. They show that if all the agents use their proposed learning approach, i.e., Generalised Toll-based Q-learning, they will converge to a system optimum. 

Mouaddib et al.\ \cite{mouaddib2007towards} develop a multi-objective multi-agent planning framework in the form of a regret-based algorithm to improve the resulting social behaviour for the considered vector-valued Dec-MDP. This framework assumes a true objective -- the regular Dec-MDP objective -- and then adds two extra objectives, i.e., the positive and negative impact an agent has on the team. This leads to social behaviour, and therefore better functioning teams. Such addition of artificial objectives to attain better policies can be seen as a form of multi-objectivisation \cite{brys2014multi}. Multi-objectivisation to improve team behaviour through social welfare is an interesting research direction that we believe needs to be further explored. 

Grandoni et al.\ \cite{grandoni2010utilitarian} assume individual utility functions with a primary objective that should be maximised, and other objectives that need to achieve at least a threshold value. This type of utility function is similar to constrained MDPs \cite{altman1999constrained} in the single-agent literature, and can be seen as a special case of multi-objective MDPs, i.e., one where the utility has this shape. They show that truthful mechanisms exist that allow finding solutions in reasonable time. 

Pla et al.\ \cite{pla2012multi} study auctions in which agents make bids that lead to a value-vector in different objectives. For such auctions the social welfare must be optimised via a mechanism that determines the payments w.r.t. the bid. These payments, together with the costs of the bids, constitute the individual utilities. They show that the social welfare function must obey three properties for a mechanism to be possible: it must be real-valued and monotonically increasing in all objectives -- as to utility functions for any multi-objective decision problem -- and it must be bijective, i.e., given the bid attributes values and the result of an evaluation function, the cost corresponding to a bid can take only one possible value. This is necessary to be able to calculate the payments in a mechanism.

Mechanisms for multi-objective decision problems are used in a variety of applications, for example: Buettner and Landes \cite{buettner2012web} apply mechanism design in order to match employers looking for temporary workers, to workers looking for contracts. Because these contracts have several aspects that may lead to utility, as hourly salary, benefits, sick pay  or overtime premiums, this is a multi-objective setting; Fard et al.\ \cite{fard2011bi} use mechanism design in for cloud work-flow management, where the agents have costs and completion time as objectives when trying to schedule tasks; and Kruse et al.\ \cite{kruse2013designing} study multi-objective airline service procurement using mechanisms. 
 
\section{Conclusions and New Horizons}
\label{sec:Future}

In this paper, we analysed multi-objective multi-agent decision problems from a utility-based perspective. Starting from the execution phase and working backwards, we derived when different solution concepts apply. We surveyed the literature on the applicable solution concepts, methods that compute such solutions, and practical applications. The taxonomy of problem settings and solution methods we propose structures this relatively new line of research from the perspective of user utility, and it is therefore our hope that this survey helps to place existing research papers in the larger multi-objective multi-agent decision problem context, and informs and helps to inspire further research. To this end, in this last section we discuss what we consider to be the key new horizons and open problems in the field of multi-objective multi-agent decision making.

\subsection{Optimisation Criteria and Solution Concepts}
In future work, it would be worthwhile to further explore the link between multi-objective optimisation criteria (ESR vs. SER) and solution concepts for MOMAS with non-linear utility functions. The body of literature on theory and experimental results is limited up until this point with respect to this topic, apart from an initial analysis conducted by R\u{adulescu} et al. \cite{radulescu2019equilibria} for multi-objective normal-form games under SER which proves by example that Nash equilibria need not exist, and that correlated equilibria can exist under certain conditions. This line of research should be extended to sequential settings (e.g., MOSGs), as well as to consider the other solution concepts discussed in Section \ref{sec:solutions}. It is also possible that not all agents in a MOMAS would choose the same optimisation criterion; it is currently not known how mixing optimisation criteria would affect the collective behaviour of MOMAS in practice. Developing stronger theoretical guarantees, as well as a better understanding of these issues using comprehensive empirical studies represents an important research direction one can pursue. 

\subsection{ESR Planning and Reinforcement Learning and SER Game Theory}
For multi-objective multi-agent decision problems, there is a large discrepancy between the game theory literature and the planning and reinforcement learning literature. The former focuses mostly on ESR settings, while the latter focuses almost exclusively on SER settings. Perhaps this is an artefact of the single-shot nature of most game-theoretic models and the sequential nature of planning and reinforcement learning models. However, as we recently noted \cite{radulescu2019equilibria}, both optimality criteria are well-motivated, as they apply to different real-world decision problem settings, and lead to vastly different theoretical results as well as practical solutions in single-shot settings with non-linear utility functions. The same argument can be made for sequential decision making settings. Therefore, we believe that analysing sequential decision problems under ESR, and game-theoretic (single-shot) models under SER, is both highly important and almost entirely unstudied. 

\subsection{Opponent Modelling and Modelling Opponent Utility}
In single-objective reinforcement learning, an agent often aims to learn a model of the other agents' behaviours and uses this model when selecting or learning best responses. In multi-objective multi-agent settings, a good and possibly even sufficient predictor for this behaviour would be the utility function of the other agents. Therefore, explicitly estimating the utility functions of the other agents in a MOMAS is likely to be important in future research. In team-utility settings, i.e., when there is only one true utility function, Zintgraf et al.\ \cite{zintgraf2018ordered} show that this utility function can be estimated effectively by posing preference queries, and using monotonicity information about the utility function. However, this assumes that there is a single user to pose such queries to, who ``owns'' the utility function. In multi-agent settings, there may be multiple utility functions, and users, that have conflicting interests. Furthermore, if they can benefit from not revealing their true preferences, they might lie. This motivates two important open questions for future research: can we design mechanisms that force agents/users to be truthful about revealing their preferences over value/return vectors? And if not, can we estimate their utility functions solely from the agent's behaviour in a multi-objective decision problem? Albrecht and Stone recently published a comprehensive survey on opponent modelling for single-objective MAS \cite{albrecht2018survey}; many of the methods they surveyed could plausibly be adapted or extended to model other agents' intentions and utilities in MOMAS.

\subsection{Closing the Loop}
From our analysis of prior works on MOMAS in Section \ref{sec:algorithmic}, it is apparent that the field to date has been quite fractured; some settings from our taxonomy (e.g. TRTU) have received much more attention than others, and a limited number of authors are currently active in the field. Consequently, there is not yet a standardised approach to identifying and completing all the steps which are necessary for a successful application of multi-objective multi-agent decision making. 

We propose the following sequence of steps: selecting an appropriate decision making model from among those listed in Section \ref{sec:Models}, identifying which setting from our taxonomy the decision making problem fits into (Section \ref{sec:execution}), defining the environment including the state and action spaces and reward and utility schemes, selecting an appropriate solution concept (Section \ref{sec:solutions}), completing the planning/learning and/or negotiation phase, executing the policies found and finally evaluating the outcome by measuring the utilities achieved.
We hope that by following these main steps, it will become easier for other researchers to apply multi-objective multi-agent decision making theory in their work.

\subsection{Interactive approaches}
In most of the survey we have assumed that there is a separate learning or planning phase first, then a policy selection and/or negotiation phase, and finally an execution phase. However, it is also possible to elicit preferences from users while planning or learning, leading to a combined planning/learning and preference-elicitation/negotiation phase. In single-agent multi-objective systems this has been studied in \cite{roijers2017interactive,roijers2018interactive}, and in cooperative game theory by \cite{igarashi2017multi}. Furthermore, the incorporation of preference information during planning in \cite{wilson2015} can also be seen in this line. This previous research however focuses either on eliciting preferences with respect to a team utility function  \cite{roijers2017interactive,roijers2018interactive,wilson2015} or individual utilities in the context of checking whether deviations from current coalitions are desired \cite{igarashi2017multi}. Parallel negotiation and learning or planning is, to our knowledge, still unexplored territory.

\subsection{Deep Multi-Objective Multi-Agent Decision Making}
Most of research discussed in this survey so far considers domains with discrete states and actions. For challenging real-world applications of MOMAS, it will be necessary to develop methods which consider continuous or high-dimensional state and action spaces. Considerable progress has been made on developing single-objective Deep RL methods for single-agent decision making. In the last couple of years, interest in Deep MORL has intensified, although primarily in single-agent settings (see e.g. \cite{abels2018dynamic,friedman2018generalizing,kallstrom2019tunable,mossalam2016multi,nguyen2018multi,reymond2019pareto,si2017multi,tajmajer2017multi,tajmajer2018modular}). 
Very recently, single-objective multi-agent RL has received considerable attention as well \cite{foerster2016learning,foerster2018counterfactual,he2016opponent,lowe2017multi,radulescu2018deep,rashid:icml18,sunehag2018value,zheng2018weighted}. 
An important next step is therefore to extend existing Deep RL methods for multi-objective multi-agent decision making settings.

\subsection{Applications and Broader Applicability}
\label{sec:application}
Now that we have identified the different settings and solution concepts which are relevant to MOMAS, significant opportunities exist to revisit problems that were initially modelled as single-objective multi-agent decision problems using a multi-objective perspective. This could provide a richer set of potential solutions for cooperative MAS using the concept of coverage sets (Section \ref{sec:coverage_sets}), or potentially improve performance by considering additional synthetic objectives of sub-tasks explicitly (a process known as multi-objectivisation \cite{brys2014multi,brys2017multi}. The possibility also exists to use MORL techniques to develop agents which may be tuned to adopt a range of different behaviours during deployment in MAS (e.g. cooperative vs. competitive), as recently demonstrated by K{\"a}llstr{\"o}m and Heintz \cite{kallstrom2019tunable}.

\end{document}